\documentclass[a4paper,fleqn,usenatbib]{mnras}
\usepackage[T1]{fontenc}
\usepackage{times,ae,aecompl}
\usepackage{graphicx}	
\usepackage{amsmath}	
\usepackage{amssymb}	
\usepackage{wasysym}
\usepackage{xcolor}
\usepackage{multirow}
\usepackage{pdfcolfoot}
\usepackage[normalem]{ulem}
\usepackage{etoolbox}
\usepackage{graphicx}	
\usepackage{bbold}

\def\be{\begin{equation}}
\def\ee{\end{equation}}
\newcommand {\kev}{{$k$-evolution}}
\newcommand{\class}{{\sc{CLASS}}}
\def\be{\begin{equation}}
\def\ee{\end{equation}}
\definecolor{bittersweet}{rgb}{1.0, 0.44, 0.37}
\definecolor{darkgreen}{rgb}{0.34, 0.55, 0.23}

\newcommand {\gev}{{\em gevolution}}
\newcommand {\kess}{$k$-essence}
\newcommand {\DE}{{\rm DE }}
\newcommand{\HH}{\mathcal H}
\newcommand{\zetaf}{\zeta}

\usepackage{scalerel,tikz}
\usetikzlibrary{svg.path}
\definecolor{orcidlogocol}{HTML}{A6CE39}
\tikzset{orcidlogo/.pic={
 \fill[orcidlogocol] svg{M256,128c0,70.7-57.3,128-128,128C57.3,256,0,198.7,0,128C0,57.3,57.3,0,128,0C198.7,0,256,57.3,256,128z};
 \fill[white] svg{M86.3,186.2H70.9V79.1h15.4v48.4V186.2z}
 svg{M108.9,79.1h41.6c39.6,0,57,28.3,57,53.6c0,27.5-21.5,53.6-56.8,53.6h-41.8V79.1z M124.3,172.4h24.5c34.9,0,42.9-26.5,42.9-39.7c0-21.5-13.7-39.7-43.7-39.7h-23.7V172.4z}
 svg{M88.7,56.8c0,5.5-4.5,10.1-10.1,10.1c-5.6,0-10.1-4.6-10.1-10.1c0-5.6,4.5-10.1,10.1-10.1C84.2,46.7,88.7,51.3,88.7,56.8z};
}}
\newcommand\orcidicon[1]{\href{https://orcid.org/#1}{\mbox{\scalerel*{
\begin{tikzpicture}[yscale=-1,transform shape]
\pic{orcidlogo};
\end{tikzpicture}
}{|}}}}

\title[Dark energy clustering signatures]{Clustering dark energy imprints on cosmological observables of the gravitational field}

\author[Farbod Hassani, Julian Adamek \& Martin Kunz]{Farbod Hassani~\orcidicon{0000-0003-2640-4460},$^{1}$\thanks{E-mail: Farbod.Hassani@unige.ch}
Julian Adamek~\orcidicon{0000-0002-0723-6740},$^{2}$\thanks{E-mail: julian.adamek@qmul.ac.uk} 
 and Martin Kunz~\orcidicon{0000-0002-3052-7394}$^{1}$\thanks{E-mail: martin.kunz@unige.ch}
\\
$^{1}$D\'epartement de Physique Th\'eorique, Universit\'e de Gen\`eve, 24 quai Ernest Ansermet, 1211 Gen\`eve 4, Switzerland \\
$^{2}$School of Physics and Astronomy, Queen Mary University of London, 327 Mile End Road, London E1~4NS, UK
\\ }
\date{Accepted XXX. Received YYY; in original form ZZZ}
\pubyear{2020}

\begin{document}
\label{firstpage}
\pagerange{\pageref{firstpage}--\pageref{lastpage}}
\maketitle
 
\begin{abstract}
We study cosmological observables on the past light cone of a fixed
observer in the context of clustering dark energy. We focus on observables that probe the gravitational field directly, namely the integrated Sachs-Wolfe and non-linear Rees-Sciama effect (ISW-RS), weak gravitational lensing, gravitational redshift and Shapiro time delay.
With our purpose-built $N$-body code ``$k$-evolution'' that tracks the coupled evolution of dark matter particles and the dark energy field, we are able to study the regime of low speed of sound $c_s$ where dark energy perturbations can become quite large.
Using ray tracing we produce
two-dimensional sky maps for each effect and we compute their angular power spectra. 
It turns out
that the ISW-RS signal is the most promising probe 
to constrain clustering dark energy properties coded in $w-c_s^2$, as the \textit{linear} clustering of dark energy would change the angular power spectrum by $\sim 30\%$ at low $\ell$ when comparing two different speeds of sound for dark energy. Weak gravitational lensing, Shapiro time-delay and gravitational redshift are less sensitive probes of clustering dark energy,
showing variations of a few percent only.
The effect of dark energy \textit{non-linearities} in all the power spectra is negligible at low $\ell$, but reaches about $2\%$ and $3\%$, respectively, in the convergence and ISW-RS angular power spectra at multipoles of a few hundred when observed at redshift $\sim 0.85$.
Future cosmological surveys achieving
percent precision measurements will allow to probe the clustering of dark energy to a high degree of confidence. 
\end{abstract}


\maketitle

\section{Introduction}

The accelerated expansion of the Universe which has been attributed to the so called ``dark energy'' component 
was first discovered using supernova Ia observations over two decades ago independently by \cite{1999ApJ...517..565P} and  \cite{1998AJ....116.1009R}. Since then the late-time accelerating expansion has been confirmed by several independent measurements, including the cosmic microwave background (CMB) \citep{Ade:2015rim, 2003ApJS..148..175S, 2016A&A...594A..13P}, large scale structure \citep{2004PhRvD..69j3501T, 2006PhRvD..74l3507T} and baryon acoustic oscillations 
\citep{2015PhRvD..92l3516A, 2007MNRAS.381.1053P}.

In the $\Lambda$CDM standard model of cosmology the cosmological constant $\Lambda$ is responsible for the late-time acceleration. Although $\Lambda$CDM is a successful model that 
fits the current data well, $\Lambda$ is a phenomenological parameter which is not theoretically well motivated and suffers from severe fundamental issues including fine-tuning of the initial energy density and the coincidence problem [see \citet{2012CRPhy..13..566M} for a review of the cosmological constant problem]. 
The theoretical issues of the cosmological constant and tensions between some parameters in various cosmological data sets \citep{2019PhRvD.100d3504H, 2019NatAs...3..891V} that have emerged over the past years have motivated cosmologists to propose a plethora of dark energy (in the form of an {additional} scalar field or ``dark'' fluid) and modified gravity models (in the form of a modification of general relativity)
to explain the late-time cosmic acceleration \citep{2012PhR...513....1C,2016ARNPS..66...95J,Koyama:2018som}. 
Since the number of viable dark energy and modified gravity models is substantial, the effective field theory {(EFT)} approach has been developed and has become popular. {It describes the theory economically in the low energy limit using  symmetries} 
and can connect the specific theory to the observation in a {straightforward} way.  

In the near future with
high precision cosmological surveys such as Euclid \citep{Laureijs:2011gra}, {the} Dark Energy Spectroscopic Instrument (DESI) \citep{1611.00036}, the Legacy Survey of Space and Time (LSST) \citep{1211.0310} and {the Square Kilometre Array (SKA)} \citep{2015aska.confE..19S} we will be able to probe the cosmological models with percent precision {into the} highly nonlinear regime. One of the main goals of these surveys is to understand the reason behind the cosmic acceleration and to probe the nature of gravity. In the case where the data prefers an alternative dark energy scenario, accurate modeling of the cosmological observables for non-standard models up to highly nonlinear {scales} is required. {A very}
accurate method to model the cosmological predictions of these theories {on such} scales is obtained using cosmological $N$-body simulations. 

Until now several $N$-body codes have been developed for a range of alternative
gravity models \citep{Oyaizu:2008tb, Schmidt:2008tn,Zhao:2010qy,2012JCAP...01..051L,Brax_2012,2012PDU.....1..162B,Puchwein:2013lza,Wyman:2013jaa,Barreira:2013eea,Li:2013tda,Llinares:2013jza,Mead:2015yca,Valogiannis:2016ane}. These codes are based on Newtonian gravity which is not {naturally suited} for considering the non-standard dark energy models{, therefore requiring a number of approximations. While these may be justified for a crude analysis, the fact that dark energy (as opposed to dark matter) is not dominated by rest-mass density appeals more to a relativistic treatment}.

{With such applications in mind, s}ome of us have developed \gev\ \citep{Adamek:2016zes}, an $N$-body code {entirely} based on General Relativity (GR). One of the main advantages of {the} \gev\ scheme is that
{it} could be extended naturally to include dark energy  or
modified gravity models without {requiring} any approximations in the dark energy {sector such as the}
quasi-static approximation as is done in the Newtonian approaches. {A} full implementation of the EFT of dark energy encompassing many dark energy and modified gravity models {is still a formidable task. As a first step toward this goal} we have recently developed {the} \kev\ code in \cite{Hassani:2019lmy} in which we have added {a} \kess\ scalar field using the EFT language. Based on \gev\ some of us have recently {also} developed $N$-body simulations for parametrised modified gravity in \cite{2003.05927}, {and \cite{Reverberi:2019bov} have implemented $f(R)$ models in \gev.}

In this article, we study the effect of $k$-essence dark energy on the cosmological observables of the gravitational field using our $N$-body codes \gev\ and \kev{. In particular, weak gravitational lensing and the ISW-RS effect probe the gravitational potential and its time derivative directly, and they can be constrained with multiple probes independently. We additionally discuss gravitational redshift and the Shapiro time delay, even though their cosmological detection will be more challenging. We study and quantify} the signatures that th{e \kess\ }model {w}ould imprint on each observable.

In Section \ref{sec:theory} we discuss the theoretical background, focusing on \kess\ in the EFT framework of dark energy, and we {present} the relevant equations that are solved in \kev\ to {evolve} the \kess\ field. Section \ref{sec:method} is devoted to the general discussion about the cosmological observables and how non-standard dark energy or modified gravity models would affect each observable. 
The cosmological parameters {of our simulations} and the way we {construct} the past light cone for a fixed observer in our $N$-body codes to make the {synthetic sky} maps {of our observables} are explained in Section \ref{sec:Nbody}.
In Section \ref{sec:numerics} we show the numerical results from \kev\ and \gev\ and we compare the results with the linear theory prediction obtained from {the} Boltzmann code \class. Our conclusions and main {take-home} points are summarised in Section \ref{sec:conclusion}.

\section{Theory} \label{sec:theory}
In this section we briefly review the essentials of the \kess~ model especially with the focus on its effective field theory description and the \kev~code, an $N$-body code recently developed to study
{the} evolution of large scale structure in the presence of {a} non-linear \kess~scalar field, see \cite{Hassani:2019lmy} for a description.

To study the evolution of the perturbations around a homogenous flat {Friedmann}
Universe, we consider the Friedmann-Lema\^{i}tre-Robertson-Walker (FLRW) metric written in the conformal Poisson gauge,
\be
ds^2 = a^2(\tau) \Big[ - e^{2 \Psi} d\tau^2 -2 B_i dx^i d\tau  + \big( e^{-2 \Phi} \delta_{ij} + h_{ij}\big)  dx^i dx^j \Big],
\ee
where $\Psi$ and $\Phi$ are {the Bardeen potentials carrying the} 
scalar perturbations of the metric, $B_i$ is the {gravitomagnetic} vector perturbation with two degrees of freedom as we have $\delta^{ij} \partial_j B_i =0$, and $h_{ij}$ is the tensor perturbation with the {gauge condition} ${\delta^{ij} h_{ij}} = 0 = \delta^{ij}\partial_i h_{jk}$ which results in two remaining degrees of freedom.

The \kess\ model was originally introduced to naturally explain the recent accelerated expansion of the Universe through the idea of a dynamical attractor solution in which this model acts as a cosmological constant at the beginning of {the} matter dominated era without any fine tuning of the parameters  \citep{ArmendarizPicon:2000dh,ArmendarizPicon:2000ah}. This model is particularly interesting as it does not rely on coincidence or anthropic {reasoning}, unlike the cosmological constant and quintessence models\footnote{The quintessence model is a canonical scalar field model, corresponding to {the special case of the} \kess\ model w{here the} kinetic term {is canonical}.} in which the energy density today is set by tuning the model parameters. 

The $k$-essence action, an action containing at most one single derivative acting on
the field, reads
\be
S_{\DE}= \int \sqrt{-g} P(X,\varphi) d^4 x ~, \label{eq:action}
\ee
where $\varphi$ is the scalar field 
and $X=-\frac{1}{2} g^{\mu \nu} \partial_{\mu} \varphi \partial_{\nu} \varphi$ is the kinetic term of the \kess~field. In general we need to {choose a specific} form for the function $P(X,\varphi)$ to solve the equations of motion for the \kess\ scalar field. 
{Since there are many possible choices, one can instead employ }{the 
{EFT} approach to model the dynamics of dark energy. EFT}, although not a fundamental theory, 
offers several advantages: First, we can express a large class of {dark energy and modified gravity (DE/MG)} models with {a} minimal number of parameters 
in a model-independent approach {and using a unified language}. Second, the {phenomenological parameters of the} effective theory can be constrain{ed} directly by cosmological observations without being specific to any DE/MG models nor to their original motivations. Third, the effective approach {allows} the theorists to carefully examine the unexplored regions of the
space of parameters which could in principle guide towards new viable models  \citep{Gubitosi:2012hu,Creminelli:2008wc}. 

{The EFT framework is a perturbative approach describing a fundamental theory to a {certain} energy/length scale. Here, our perturbative expansion is performed in terms of the scalar field. Like the metric perturbations, we assume that the scalar field perturbations remain small over the scales of interest (i.e.\ cosmological scales). 
{The non-linear regime remains accessible to EFT as long as perturbation theory remains valid} [see \citet{Frusciante:2017nfr, Cusin:2017mzw} for a discussion about the non-linear extension of the EFT of DE framework].
For example, for the case of $k$-essence theory, we have shown in \citep{Hassani:2019lmy} that based on our EFT scheme, highly non-linear scalar field {configurations} are allowed to form while still respecting the weak field approximation. {T}he main difference between the fundamental theory and {its EFT description} is that 
the scalar field can become {arbitrarily} large in the 
former while it has to remain perturbatively small in the latter. The upshot is that the EFT framework allows one to study the infrared phenomenology without having to specify the ultraviolet completion of the theory in the first place.}

In \kev\ \citep{Hassani:2019lmy}{, which is an extension of \gev\ } 
\citep{Adamek:2015eda,Adamek:2016zes} in which we have implemented {the} \kess\ model as a dark energy sector, we use the EFT approach to write down the equations of motion parametrised with the equation of state $w$ and the {squared} speed of sound $c_s^2$.
They are related to the general action of Eq.~(\ref{eq:action}) as
\be
w=\frac{P}{2 X P_{, X}-P}\,, \qquad c_{s}^{2}=\frac{P_{, X}}{2 X P_{, X X}+P_{, X}}\,, \label{wcs2_sets}
\ee
where $P_{, X} \doteq \partial P / \partial X$ {and one has to evaluate the expressions on the background solution of $\varphi$} \citep{ArmendarizPicon:1999rj, Bonvin:2006vc}. {These phenomenological parameters completely characterise the low-energy limit of the theory. Of course, $w$ and $c_s^2$ do not uniquely fix the function $P(X,\varphi)$ and in this work we will not consider the question of finding suitable choices. Explicit examples have been studied {for example} in \citet{Chiba:1999ka,ArmendarizPicon:1999rj,Malquarti:2003nn,Wei:2004rw,Rendall:2005fv,ArmendarizPicon:2000ah,dePutter:2007ny,Silverstein:2003hf,Alishahiha:2004eh,Fang:2006yh,Kang:2007vs}.}

The scalar field evolution, keeping only linear terms of {the} scalar field and its time derivative, is given by
\begin{align}
 \pi '&  = \zetaf - \HH \pi + \Psi \;, \label{zeta_eq1_lin} \\
 \zetaf' & =  3w \HH \zetaf -  3 c_s^2 \left(\HH^2 \pi - \HH \Psi - \HH' \pi - \Phi' \right) +c_s^2 \nabla^2 \pi\;, \label{zeta_eq2_lin} 
\end{align}
where $\pi$ is the scalar field perturbation around its background value, $\zetaf$ is written in terms of $\pi$ and $\pi'$, {$\nabla^2 = \delta^{ij} \partial_i \partial_j$, a prime $\prime$} denotes the time derivative with respect to the conformal time, and $\HH$ is the conformal Hubble function.
The linear stress energy tensor reads
\be
\label{SEcomponents_lin}
\begin{split}
 T_0^0 &= - \rho  + \frac{\rho + p}{c_s^2}  \Big( 3c_s^2 \HH \pi -\zetaf   \Big) \;, \\
 T^0_i & = - (\rho+p) \partial _i \pi  \; ,\\ 
T_{j}^{i}&  =  p \delta_{j}^{i}  - (\rho+p)  \Big( 3 w  \HH \pi -\zetaf  \Big) \delta_{j}^{i}  \;.
\end{split}
\ee
It is worth {noting} that in connection with the EFT language, with the notation used in \cite{Gleyzes:2013ooa}, apart from the background parameters $\Lambda(t)$ and $c(t)$ the only remaining parameter in EFT for $k$-essence theory is $M_2^4(t)$. In the {alternative} EFT notation using $\alpha_i(t)$ {explained} in \cite{Bellini:2014fua}, $\alpha_{K}$ is the only non-zero parameter. 

{The detection of the binary neutron star merger GW170817 and its electromagnetic counterpart \citep{Monitor:2017mdv} implies that gravitational waves propagate at the speed of light with uncertainties of about $5 \times 10^{-16}$. This result puts strong constraints on  dark energy and modified gravity theories 
\citep{Creminelli:2017sry, Ezquiaga:2017ekz, Copeland:2018yuh, Sakstein:2017xjx, Baker:2017hug}. 
{The speed of gravitational waves is modified in any theory that features a non-vanishing $\alpha_T$ parameter, which is now observationally strongly disfavoured. However, this constraint does not affect}
the $k$-essence model because, as we just pointed out, only $\alpha_K$ is non-zero in this model.} 

As discussed in \cite{Hassani:2019lmy}, although we keep only linear terms in the scalar field dynamics and we drop higher order self-interactions of $\pi$, the scalar field does cluster and form non-linear scalar field structures in \kev\ {because} it is sourced by non-linearities in matter {through gravitational coupling}. {This is the crucial improvement over the treatment of dark energy in \gev, which uses the transfer functions from linear theory to keep track of perturbations in the dark energy fluid similar to how it is done in \citet{Dakin:2019vnj} [see also \citet{Brando:2020ouk} for a more comprehensive extension of this framework to EFT].}

Einstein's equations for the evolution of the scalar perturbations in the weak field regime \citep{Adamek:2017uiq} read
\begin{multline}
 (1+2 \Phi) \nabla^2 \Phi -3 \mathcal{H} \Phi^{\prime}-3 \mathcal{H}^{2}{\Psi}-\frac{1}{2} \delta^{i j} \partial_i \Phi \partial_j \Phi \\   =-4 \pi G a^{2} \delta T_{0}^{0} \;,
\end{multline}
\begin{multline}
\nabla^4 {\left(\Phi-\Psi\right)} -\left(3 \delta^{i k} \delta^{j l}  \frac{\partial^{2}}{\partial x^{k} \partial x^{l}}-\delta^{i j} \nabla^2\right) {\partial_i}\Phi {\partial_j}\Phi  \\   = 4 \pi G a^{2}\left(3 \delta^{i k} \frac{\partial^{2}}{\partial x^{j} \partial x^{k}}-\delta_{j}^{i} \nabla^2 \right) T_{i}^{j} \;,
\end{multline}
where the stress tensor $T_{\mu}^{\nu}$ includes all the relevant species{, i.e.\ }matter, dark energy and radiation. 

In \cite{Hassani:2019lmy}, we introduce $k$-evolution and the full non-linear equations for the \kess~model written using  the EFT action. We study the dark energy clustering and its impact on the large scale structure of the Universe. Moreover, we discuss that the scalar dark energy does not lead to significantly larger vector and tensor perturbations than $\Lambda$CDM. Thus one can safely neglect the vector and tensor perturbation in this theory.
In \cite{Hassani:2019wed}  we quantify the non-linear effects from \kess~dark energy through the effective parameter $\mu(k,z)$ that encodes the contribution of a dark energy sector to the Poisson equation {(see below)}. We also show that for the $k$-essence model the difference between the two potentials $ \Phi -\Psi$ and short-wave corrections appearing as higher order terms in the Poisson equation can be safely neglected. Moreover, in \cite{Hansen:2019juz}, we study the effect of $k$-essence dark energy as well as some other modified gravity theories on the turn-around radius --- the radius {at} which the inward velocity due to the gravitational attraction and outward velocity due to the expansion of the Universe cancel each other --- in galaxy clusters.

\begin{figure*}
\begin{minipage}{\textwidth}
\centering
\includegraphics[width=.328\textwidth]{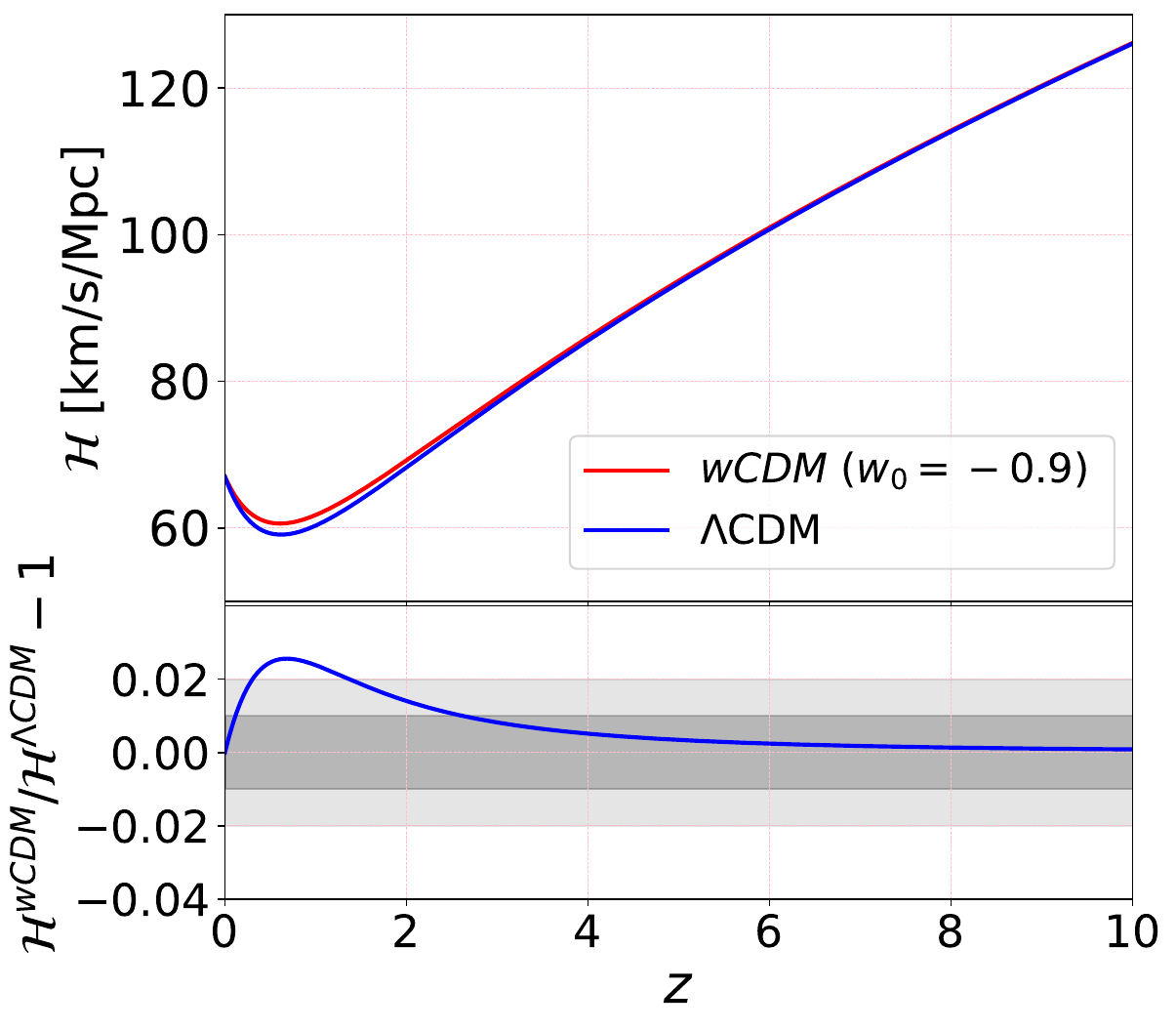}\hfill
\includegraphics[width=.33\textwidth]{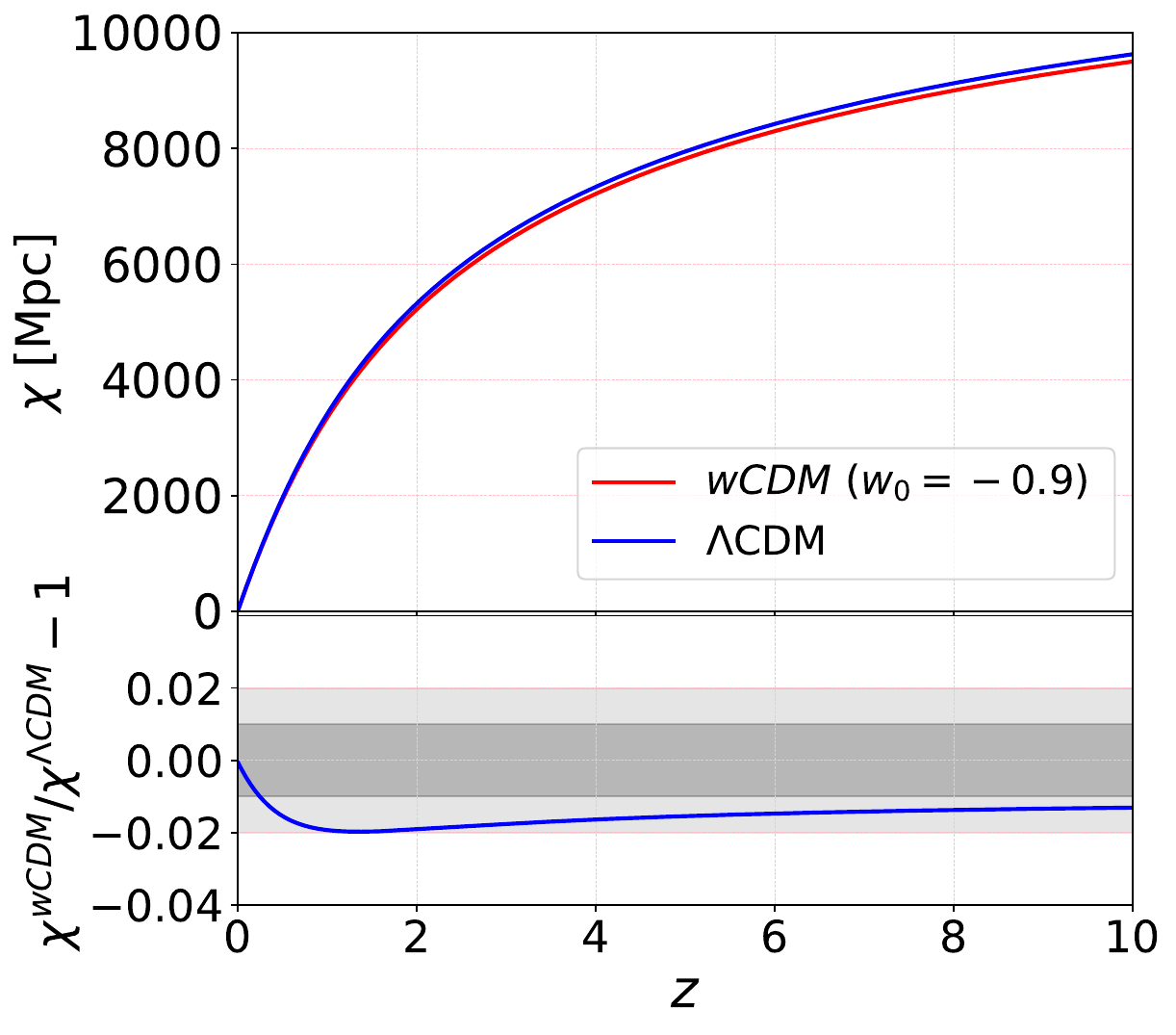}\hfill
\includegraphics[width=.324\textwidth]{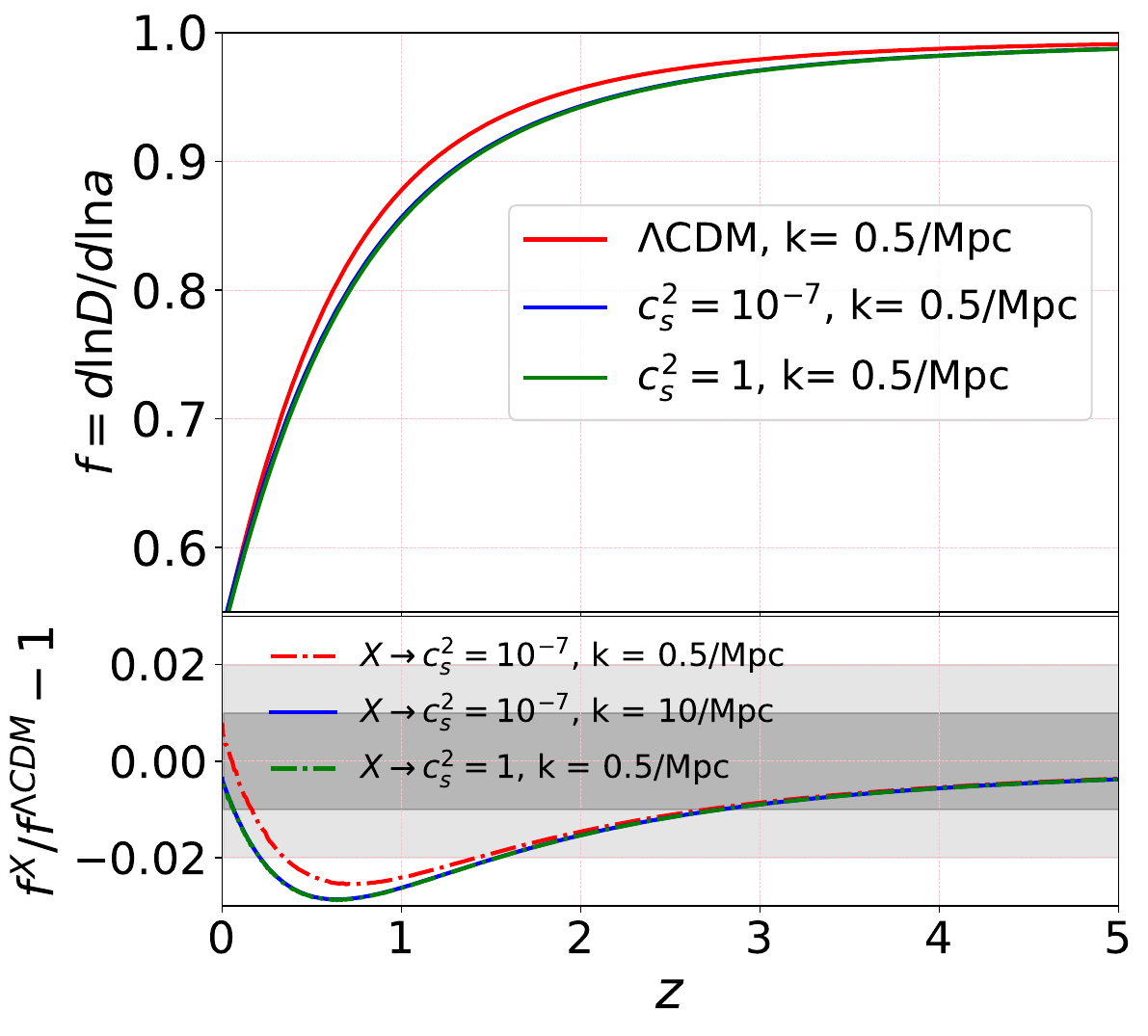}
\caption{ {\bf{\emph{Left:}} }The conformal Hubble parameter as a function of redshift for $\Lambda$CDM and $w$CDM with $w_0=-0.9$. In the bottom {panel we show} the relative difference. 
{As expected,}
the effect of dark energy goes away at high redshifts. 
{\bf{\emph{Centre:} }}The comoving distance as a function of redshift for {the two} cosmologies is plotted. In the bottom {panel again} the relative difference 
 is shown. The relative difference 
 {asymptotes} to a constant at high redshifts because the comoving distance is an integrated quantity.
{\bf{\emph{Right:}}} In the top {panel}, the logarithmic growth rate $f$ as a function of redshift for a fixed wavenumber is plotted. In the bottom the relative difference between different scenarios {is shown with respect to} $\Lambda$CDM.
{While in the latter case the growth rate is scale independent (neglecting corrections due to neutrinos), in a clustering dark energy model t}he behaviour of a mode depends on whether it is inside or outside {of the} sound horizon. We compare three cases with $\Lambda$CDM, namely {the} mode $k = 0.5 $ Mpc$^{-1}$ {with} $c_s^2=10^{-7}$, $k=10$  Mpc$^{-1}$ {with} $c_s^2=10^{-7}$ and $k = 0.5 $ Mpc$^{-1}$ {with} $c_s^2=1$. 
{Only the first case corresponds to a mode outside the sound horizon, while the other two are well inside.}
Thus we see the same behaviour for $k=10$ Mpc$^{-1}$ {with} $c_s^2=10^{-7}$ and $k = 0.5 $ Mpc$^{-1}$ {with} $c_s^2=1$.
}
\label{fig:BG}
 \end{minipage}
 \end{figure*}

{It is sometimes useful to parametrise the effect that} clustering dark energy {has on the} gravitational potential {phenomenologically through a modified Poisson equation,} 
\be
-k^{2} \Phi=4 \pi G_{N} a^{2} \mu(k, z) \sum_{X} \bar{\rho}_{X} \Delta_{X}\;, \label{eq:Hamiltonian}
\ee
where $\Delta$ is the comoving density contrast {and the species $X$ do not include the dark energy field}. {Denoting in addition} the ratio between the two {Bardeen} potentials as
\be
\eta (k,z) = \frac{\Psi(k,z)}{\Phi(k,z)} \label{eq:slip}\;,
\ee
{one obtains a phenomenological classification of DE/MG models through two generic functions of time and scale, with $\Lambda$CDM predicting $\eta = \mu = 1$ everywhere \citep[e.g.][]{Ade:2015rim,1910.09273}.}
As discussed in \citet{Hassani:2019wed}, for the case where $k$-essence plays the role of dark energy, $\eta(k,z) \approx 1$ and $\mu(k,z)$ is fitted well with {a} $\tanh$ function where the amplitude and shape of the function depends on the speed of sound and equation of state.
{In \citep{Hassani:2019wed} we studied the difference between $\mu$ from linear and non-linear treatment, and we have shown 
that for high speed of sound 
one would get similar $\mu$ {in both cases,} 
but for low speed of sound there is an effect coming from the scalar field non-linearities which should be taken into account. Here, the speed of sound is considered ``low'' if the corresponding sound horizon lies {at} a non-linear scale.  }

For the sake of completeness, we show the effect {of} $k$-essence, expressed in the form of $w-c_s^2$ in the parameter space, on the background evolution and the growth of matter density perturbation. The effect of $k$-essence appears on the background evolution via {the} equation of state {but} independent of the speed of sound of dark energy. {W}e show the conformal Hubble parameter $\HH$ as well as {the} redshift-distance relation in Fig.~\ref{fig:BG} in the left and middle panel for $w$CDM with $w=-0.9$ and $\Lambda$CDM, which are the cases we consider in this article.  We also compare the logarithmic growth rate $f=d \ln \Delta_m / d \ln a$ in the right panel {of Fig.~\ref{fig:BG}} for different scenarios to see the impact of  different {choices} of $w$ {and} $c_s^2$ on the growth of structures. The logarithmic growth rate
{provides a}
useful way to distinguish between DE/MG  gravity and $\Lambda$CDM 
\citep{Kunz:2006ca,Amendola:2016saw,Dossett:2013npa,1910.09273}. 

{Since the logarithmic growth rate is in general scale dependent, we compute it for different scales by studying the evolution of the comoving matter density contrast $\Delta_m$, which we define for this purpose through $\Omega_m \Delta_m = \Omega_b \Delta_b + \Omega_\text{cdm} \Delta_\text{cdm}$. We compute}
$\Delta_m$ {with} the linear Boltzmann code \class\ and 
look at the evolution {for} wavenumbers $k = 0.5 $ Mpc$^{-1}$ and $k=10$  Mpc$^{-1}$ for the model with $c_s^2=10^{-7}$ which corresponds respectively to {a} mode well {out}side and {a mode} well {in}side {the} sound horizon. We compare the logarithmic growth rate from the two modes for $c_s^2=10^{-7}$, and {in addition} $k = 0.5 $ Mpc$^{-1}$ for $c_s^2=1$, with $\Lambda$CDM in the bottom {right} panel {of} Fig.~\ref{fig:BG}. As expected, for the mode $k = 0.5 $ Mpc$^{-1}$ in {the case where} $c_s^2=1$, as it is well inside {the} sound horizon, the result matches with the mode evolution of $k=10$  Mpc$^{-1}$ in {the case where} $c_s^2=10^{-7}$. The mode $k = 0.5 $ Mpc$^{-1}$ {has a different evolution in the case where} $c_s^2=10^{-7}$ {because it is outside the sound horizon and the perturbations in the dark energy field are therefore not suppressed}.

\section{{Observables}} \label{sec:method}
Solving the null geodesic equation (${k^{\mu} \nabla_\mu k_{\nu}} = 0$) for a light ray with {wavevector} $k^\mu$ in a {perturbed} FLRW universe results in a change in position and energy of the light beam emitted from the source. 
The apparent comoving 3D position of the source reads \citep{2011PhRvD..84f3505B, 2011PhRvD..84d3516C, 2009PhRvD..80h3514Y, Breton:2018wzk,Adamek:2019vko}
\begin{multline}
{\boldsymbol{s}} = \chi_{s} \boldsymbol{n}
+\!\!\!\!\!\!\underbrace{\frac{1}{H_s(\bar z)} \delta z \, \boldsymbol{n} }_{\text {redshift perturbation}}\!\!\!\!\!\! - \underbrace{\int_{0}^{\chi_s}\left(\chi_s-\chi \right) \nabla_{\perp}(\Phi+\Psi)\mathrm{d} \chi }_{\text {weak gravitational lensing}} \\
-  \underbrace{\boldsymbol{n} \int_{0}^{\chi_s}(\Phi+\Psi) \mathrm{d} \chi}_{\text {Shapiro time delay}}\;,
 \label{eq:position}
\end{multline}
where {$\boldsymbol{s}$ is the observed position in redshift space, and} $\boldsymbol{n}$ {and $\chi_s$} are respectively the {unperturbed direction vector and unperturbed comoving distance to the source. Furthermore,} 
$\nabla_{\perp} {\equiv (\mathbb{1} - \boldsymbol{n}\otimes\boldsymbol{n})\nabla}$ is the 
{projected gradient perpendicular to the line of sight,}
$\chi$ is the comoving distance and $H_s(\bar z)$ is the Hubble {rate} at the unperturbed redshift $\bar{z}$ of the source. The first term in Eq.~(\ref{eq:position}) corresponds to the unperturbed position, the second term corresponds to the redshift perturbation which is computed in Eq.~(\ref{eq:delta_z}), the third term 
(weak gravitational lensing) yields an angular deflection of the source position on the 2D sphere{, and the last term changes the radial position due to the Shapiro time delay.}

Using the time component of geodesic equation {one obtains t}he redshift perturbation  ${\delta z =} z  -\bar{z}$ as
\begin{align}
\delta z = \left(1 + \bar{z}\right) \biggl[ \underbrace{\boldsymbol{n} \cdot \left(\boldsymbol{v}_s - \boldsymbol{v}_o\right)}_{\text {Doppler }}
  + \!\!\underbrace{\Psi_{o}-\Psi_{s}}_{\substack{\text{gravitational}\\ \text{redshift}}}\!\!
 -\underbrace{\int_{0}^{\chi_s} \!\frac{\partial( \Psi + \Phi)}{\partial \tau} d \chi}_{\text {ISW-RS effect }}\bigg]\;. \label{eq:delta_z}
\end{align}
As a result of Eq.~(\ref{eq:position}) and Eq.~(\ref{eq:delta_z}), the light ray along its path is deflected, redshifted and delayed due to the different physical phenomena, namely weak gravitational lensing, integrated Sachs-Wolfe and Rees-Sciama (ISW-RS) effect {and} Shapiro time delay. {In addition, there are two local (non-integrated) contributions to the redshift: the {relativistic} Doppler {effect (due to the peculiar velocities $\boldsymbol{v}_s$ and $\boldsymbol{v}_o$ of the source and observer, respectively)} and {the} gravitational redshift (ordinary Sachs-Wolfe effect).} Among these, the weak gravitation lensing, ISW-RS, Shapiro time delay 
and gravitational redshift
{all depend directly on the configuration of the gravitational field. If they can}
be observed independently {they can} be used as a probe of DE/MG. In the current {work} we are going to {look at} each {individual} aforementioned effect to study the impact of $k$-essence clustering dark energy. It {is} important to {point out that} the Doppler term is also expected to be {a powerful} probe of dark energy and is worth to be studied in detail. However, in this paper we only focus on the physical effects coming from the metric perturbations $\Psi$ and $\Phi$. {These have the advantage that they can be observed (as integrated effects) even in places where there are no visible sources, like e.g.\ in voids which are naturally expected to be dominated by dark energy.}

In the following subsections we shall introduce and derive the relevant expressions for each effect separately and we will discuss why each of these observables could be potentially used to put constraint on DE/MG models, especially clustering dark energy.
\subsection{\label{sec:wl}Weak gravitational lensing}
The light from distant objects {is} deflected due to inhomogeneities as {it travels} through the inter{vening} large scale structure in the Universe. The deflection {angles} are usually small at cosmological scales and this phenomen{on} is called weak gravitational lensing. {As mentioned already, w}eak gravitational lensing directly probes the distribution {of matter and energy}, including dark matter {and dark energy}, which makes it a unique tool to constrain the cosmological parameters \citep{Bartelmann:1999yn, Hikage:2018qbn,Ade:2013tyw, Lewis:2006fu, Hassani:2015zat, Refregier:2003ct}.
Weak gravitational lensing can be observed through the statistics of cosmic shear and cosmic convergence, as we will summarise briefly in the following.

Cosmic shear {refers to the change of the} ellipticit{ies} of galaxies {observed on the far side of the gravitational lens}. The cosmic convergence, {on the other hand}, magnifies/demagnifies the sizes and magnitudes of the {same} galaxies \citep{Alsing:2014fya, Mandelbaum:2017jpr}.
The first detection of cosmic shear have been done about 20 years ago \citep{2000A&A...358...30V, 2000MNRAS.318..625B}, while cosmic convergence {was measured}
for the first time in 2011 \citep{Schmidt2011}.
Since then the precision in cosmic shear and cosmic convergence measurements has been improved and these two have become {some} of the most promising probes of dark energy and modified gravity \citep{PhysRevLett.91.141302, SpurioMancini2018, Hannestad2006, Amendola2008}.

In \cite{Adamek:2018rru}, {some of us have implemented a ray-tracing method to analyse relativistic $N$-body simulations performed with the code \gev. In this method one solves} 
the optical equations 
in the scalar {sector of gravity without any approximation and additionally keeps track of} 
the frame dragging to first order. 
{Weak-lensing convergence and shear obtained with this numerical method were studied in more detail}
in \cite{Lepori:2020ifz}. 
{While other approaches often try to reconstruct the signal from the mass distribution, our method works with the metric perturbations directly and is therefore more robust once we consider DE/MG like in the present work. To this end w}e have recently implemented the {same} light-cone analysis and ray-tracing method in \kev. 
{Here, however, we compute each effect only to first order in $\Phi$ and $\Psi$, and we neglect the frame dragging. For the purpose of discussing angular power spectra this is sufficient, as nonperturbative effects (in the ray tracing\footnote{We remind the reader that $\Phi$ and $\Psi$ themselves represent nonperturbative solutions obtained with full simulations.}) would only enter at a detectable level for very high multipoles that are not resolved in our maps.}

In the first order weak gravitational lensing
{formalism one introduces}
the lensing potential {$\psi(\boldsymbol{n}, z)$} defined as
\be
\psi(\boldsymbol{n}, z) \equiv-\int_{0}^{\chi_s} d {\chi} \frac{\chi_s-\chi}{\chi_s \chi}(\Phi+\Psi)\;, \label{eq:lensingphi}
\ee
where 
{the redshift $z$ and the source distance $\chi_s$ are related through the background distance-redshift relation, and the integration is carried out in direction $\boldsymbol{n}$ using the Born approximation.}

Using the {gradient $\hat{\nabla}^a$} on the 2D sphere we can calculate the deflection angle as {$\alpha^a = \theta^a_o - \theta^a_s = \hat{\nabla}^a \psi$}, which results in the {lensing} term in Eq.~\eqref{eq:position}. The Jacobi map, which is a map between the {unperturbed} source angular positions $\theta_s$ {and} the observed angular positions $\theta_o$, i.e. {$\mathcal{A}_{ab} = {\partial \theta^a_o}/{\partial \theta^b_s}$,}  contain{s} the full information {for} weak gravitational lensing {and reads}
\be
\mathcal{A}=\left(\begin{array}{cc}1-\kappa -\gamma_{1} & {\omega}-\gamma_{2} \\ -{\omega}-\gamma_{2} & 1-\kappa+\gamma_{1}\end{array}\right)\,,
\ee
which at leading order is rewritten in terms of the {gradients} of the lensing potential,
\be
\mathcal{A}=\mathbb{1}+\left({\hat{\nabla}_a \hat{\nabla}_b \psi}\right) \,. \label{eq:amplificationmatrix}
\ee
{Note that this also implies $\omega = 0$ at leading order. W}e can extract {the} convergence $\kappa$ and {the} complex shear from the lensing potential as
\begin{align}
& \kappa  =-\frac{1}{2} {\hat{\nabla}^2 \psi\,,} \label{eq:kappa} \\ &
\gamma \equiv \gamma_{1}+i \gamma_{2} =-\frac{1}{2}\left( {\hat{\nabla}_{1} ^2 -  \hat{\nabla}_{2}^2}\right) {\psi} - \mathrm{i}  {\hat{\nabla}_{1} \hat{\nabla}_{2} \psi} \label{eq:gamma} 
\end{align}
The relation between {a non-}perturbative {geometrical description} and the{se} first-order quantities is discussed in detail in \citet{Lepori:2020ifz}. 

The lensing potential, convergence and complex shear 
as scalar functions on {the} 2D sphere may be expanded in {the basis} of spherical harmonics {$Y_{\ell m}(\boldsymbol{n})$},
 \be
 X(\boldsymbol{n})=\sum_{\ell m} X_{\ell m} Y_{\ell m}(\boldsymbol{n})\,.
 \ee
Assuming {that these functions obey} statistical isotropy, {the two-point function of the expansion coefficients $X_{\ell m}$ is diagonal in $\ell$ and $m$ and we can define the angular power spectrum $C_\ell$ as,}
\be
\left\langle X_{\ell m} X_{\ell^{\prime} m^{\prime}}\right\rangle \equiv\delta_{\ell \ell^{\prime}} \delta_{m m^{\prime}} C_{\ell}^{X} \, . \label{eq:harmonics}
\ee
{Our code uses directly Eq.\ (\ref{eq:lensingphi}) to {construct} the lensing map and then computes the angular power spectrum from the map. We can however also compute an expression for the {weak-}lensing $C_\ell^\kappa$ by considering the two-point function of {the integrand of Eq.} (\ref{eq:lensingphi}) {and using Eq.~(\ref{eq:kappa}) in {harmonic} space,}
\begin{multline}
{C_\ell^\kappa = 4 \pi^2 \ell^2 \left(\ell + 1\right)^2 \!\int_0^\infty\! k^2 dk \int_0^{\chi_s}\! d\chi\frac{\chi_s-\chi}{\chi_s \chi}  \int_0^{\chi_s}\! d\chi' \frac{\chi_s-\chi'}{\chi_s \chi'} } \\
{\times\left[1+\eta(k, \chi)\right] \left[1+\eta(k, \chi')\right] j_\ell(k\chi) j_\ell(k\chi') P_\Phi(k,\chi,\chi')\,,}\label{eq:Clkappa}
\end{multline}
where $P_{\Phi}(k, \chi{, \chi'})$  is the {unequal-time correlator of the} gravitational potential {in Fourier space} defined by
\be
\left\langle\Phi(\boldsymbol{k} , \chi) \Phi^{*}\left(\boldsymbol{k}^{\prime} , \chi' \right)\right\rangle=(2 \pi)^{3} \delta_{\mathrm{D}}\left(\boldsymbol{k}-\boldsymbol{k}^{\prime}\right) P_{\Phi}\left(k , \chi,\chi' \right)\,.
\ee
Moreover, according to Eq.~(\ref{eq:Hamiltonian}) the gravitational potential {unequal-time correlator} may be written based on {the one of} matter as follows,
\begin{multline}
 P_{\Phi}\left(k , \chi{, \chi'}\right)= k^{-4} \left(4 \pi G_{N} \bar{\rho}_{m}^0 \right)^{2}\mu(k, {\chi}) \mu(k, {\chi'}) \\ {\times \left[1+\bar{z}(\chi)\right] \left[1+\bar{z}(\chi')\right]} P_{\Delta_m}\!\left(k , \chi{, \chi'} \right)\,.\label{eq:Ppot}	
\end{multline}
{In the last few equations, we use the shorthand $\mu(k,\chi)$ for $\mu\left(k,z(\chi)\right)$ and similar for $\eta$, $P_\Phi$ and so on.}

The lensing signal, as a result, {responds} to the DE/MG models in {multiple} ways{: at the background level through a change in the distance-redshift relation, and at the level of perturbations through a modified growth and through the modifications} 
encoded in the $\mu(k,z)$ and $\eta(k,z)$ parameters \citep{SpurioMancini2018, Takahashi2017}, {specifically through the combination $\Sigma = \mu(1+\eta)/2$ that describes the modification of the lensing potential \citep{Amendola2008}.}
In Sec.~\ref{sec:numerics} we will {compute} the lensing signal for two fixed {source} redshifts, namely $z=0.85$ and $z=3.3$, in $N$-body simulations to study the effect{s} of dark energy clustering and {of the} expansion history on the lensing.

It {is} also worth mentioning that at leading order in the absence of systematics and shape noise \citep{Kohlinger:2017sxk} the convergence angular power spectrum contains the full lensing information. Decomposing the shear into rotationally-invariant E and B components 
{one can show from Eq.~(\ref{eq:amplificationmatrix}) that} {\citep{10.1093/mnras/stt1352}
\be
\begin{aligned} C^{\gamma_\mathrm{E}}_\ell &= \frac{1}{\ell^{2}(\ell+1)^{2}} \frac{(\ell+2) !}{(\ell-2) !} C^{\kappa}_\ell\,, \\ C^{\gamma_\mathrm{B}}_\ell &= \frac{1}{\ell^{2}(\ell+1)^{2}} \frac{(\ell+2) !}{(\ell-2) !} C^{\omega}_\ell=0\,. \end{aligned}
\ee
{In} the following sections we {therefore} only discuss the convergence power spectra.

\subsection{Integrated Sachs-Wolfe and non-linear Rees-Sciama effect}
\label{sec:ISW_RS}
{If the light rays from source galaxies traverse a time-dependent gravitational potential then in general their energy will change \citep{Sachs:1967er}. Specifically, the light is redshifted for rays passing through a growing potential well, and blueshifted if the potential well is decaying.}
This effect is known as late integrated Sachs-Wolfe effect (late ISW) and is a powerful probe of dark energy at low multipoles.

{During} matter domination and {in} linear perturbation theory the gravitational potential wells remain constant in time and as a result the photons do not gain or lose energy along their trajectories {after accounting for the expansion of the background}. {But when dark energy becomes important and the expansion rate of the Universe deviates from the matter-dominated behaviour,} the gravitational potential decays at {linear} scales{. As a result,} photons gain energy and are blueshifted as they travel through {over-dense regions while they lose energy and are redshifted as they travel through voids} \citep{Carbone:2016nzj, Ade:2015rim}. {In this way,} the late ISW effect
induces {additional} anisotropies {in the power spectrum of the cosmic microwave background (CMB) radiation, primarily at large scales. These anisotropies are correlated with the large-scale structure as they are due to evolving gravitational potential wells.}

In addition to {this effect in linear perturbation theory}, the perturbations {on small scales and at late times evolve non-linearly}. The non-linear large scale structure of the Universe {induces} additional energy changes in the light rays as they pass through these structures. This {so-}called Rees-Sciama effect  \citep{1968Natur.217..511R} enhances the ISW effect in under-dense regions 
and decreases it in the over-dense regions as the non-linear growth of structure acts opposed to dark energy \citep{2010MNRAS.407..201C}.

The {linear} late ISW and non-linear Rees-Sciama effects {(abbreviated as ISW-RS when combined)} are sensitive to the background evolution and growth of structures at late times, when the dark energy dominates over other components \citep{Cabass2015,Adamek:2019vko,Beck2018,Khosravi:2015boa}.

Direct measurements of the ISW-RS signal from CMB data is demanding, and as a result 
{it is}
detected indirectly, either by cross-correlating large scale structure data and CMB maps \citep{Scranton:2003in,Seljak1996,Francis2010,Peiris2000,Planck2016ISW}  or by stacking clusters and voids {to enhance the signal} \citep{Granett2008,Planck2016ISW,b309f8a96352444bb8f144aa6ab78192,Cai2014}. The effect is also detected through the ISW-lensing bispectrum using Planck data \citep{Planck2014}.

{According to} Eq.~\eqref{eq:delta_z}, 
{the term}
responsible for the ISW-RS effect 
yields {a} change in CMB temperature, 
\be
\Theta (\boldsymbol{n}, z) \equiv \frac{\Delta T}{\bar{T}} = -\frac{\delta z}{1+\bar{z}}  =  \int_0^{\chi_{\mathrm{s}}} \frac{\partial (\Psi +\Phi)}{\partial \tau} \; d \chi\,,\label{eq:isw}
\ee
{where $\chi_s$ is the distance to the last-scattering surface for the CMB. In the presence of}
DE/MG the gravitational potentials $\Phi$ and $\Psi$ are modified as a result of  Eq.~\eqref{eq:Hamiltonian} and Eq.~\eqref{eq:slip} for the gravitational slip and clustering parameters {$\eta(k,z)$,} $\mu(k,z)$. It {is} also interesting to note that the{se modifications are} projected {in a different way for the} ISW-RS signal {and the} lensing signal, and thus these two could probe DE/MG in independent way{s}.
Following the discussion in the previous subsection, $\Theta ({\boldsymbol{n}}, z)$ may be expanded in terms of spherical harmonics and one can compute the ISW-RS angular power spectrum. Taking the time derivative of {the} Hamiltonian constraint (the modified Poisson equation) \eqref{eq:Hamiltonian}{, and replacing $\Psi$ through \eqref{eq:slip},} result{s} in a constraint equation for $\partial \left(\Phi + \Psi\right)/\partial\tau$ which in general is a function of $\mu(k,z)$, $\eta(k,z)$ and their time derivatives{:
\begin{multline}
\frac{\partial(\Phi + \Psi)}{\partial\tau} = 4 \pi G_N a^2 \bar{\rho}_m \Delta_m \frac{H(z)}{k^2} \biggl[\mu(k,z) \frac{\partial\eta(k,z)}{\partial z}\biggr. \\ \biggl.-\left(1 + \eta(k,z)\right)\left(\mu(k,z)\frac{f(k,z)-1}{1+z}-\frac{\partial\mu(k,z)}{\partial z}\right)\biggr]
\end{multline}
Writing this as $\partial \left(\Phi + \Psi\right)/\partial\tau = g(k,z) \Delta_m$ we find
\begin{multline}
C_\ell^\Theta = 16\pi^2 \int_0^\infty\! k^2 dk \int_0^{\chi_s}\! d\chi \int_0^{\chi_s}\! d\chi' j_\ell (k \chi) j_\ell (k \chi') \\\times g(k,\chi) g(k,\chi') P_{\Delta_m} (k, \chi, \chi')\,.
\end{multline}
}

{In} this article we use 
ray tracing to compute {the} ISW-RS signal {according to Eq.~\eqref{eq:isw}} by integrating {along the} past light cone to {the} source redshift{s $z=0.85$ and $z=3.3$ as in the previous section. This can be seen either as a direct contribution to the observed redshift of the sources, or as a fractional contribution to the CMB temperature anisotropy that captures the part of the signal that is generated by the structure out to that distance and that could therefore be constrained through a cross-correlation of the CMB with large-scale structure}. We use the ISW-RS angular power spectra in Sec.~\ref{sec:numerics} to measure the response of this signal to different $k$-essence scalar field scenarios.

\subsection{Gravitational redshift} \label{sec:gravitational_redshift}
The light rays emitted from source galaxies in the potential well of galaxy clusters and dark matter halos are expected to be redshifted/blueshifted due to the difference in gravitational potential between the source galaxy and the observer. This effect is known as gravitational redshift \citep{1995A&A...301....6C}, {or} as ordinary Sachs-Wolfe effect in {the context of} CMB physics \citep{Durrer:2001gq}, and is the second term in the full expression written in Eq.~(\ref{eq:delta_z}),
\be
\delta z_{\text{grav}} = {(1+\bar{z})}  \left( \Psi_o - \Psi_s \right) \, . \label{eq:gr_redshift}
\ee
For typical cluster masses (of order $\sim 10^{14} M_\odot$) the gravitational redshift is estimated to be two orders of magnitude smaller than the Doppler shift [the first term in Eq.~(\ref{eq:delta_z})] coming from the random motion of source galaxies \citep{1995A&A...301....6C}. The technique to extract the gravitational redshift signal from other dominant signals depends on the fact that the Doppler shift results in a symmetric {dispersion in the redshift-space distribution}, while the gravitational redshift {changes the mean of the} distribution \citep{Broadhurst2000,Kim2004}. In \cite{Wojtak:2011ia} the first measurement of gravitational redshift of light coming from galaxies in clusters was carried out by stacking 7800 clusters from the SDSS survey in redshift space. The signal detection was used to rule out models avoiding the presence of dark matter and also to show the consistency of the results with the predictions of general relativity. The gravitational redshift signal was detected [e.g.\ {in} \citet{PhysRevLett.114.071103,Jimeno:2014xma}] on scales of {a} few Mpc around galaxy clusters and recently in  elliptical galaxies \citep{zhu2019gravitational} using spectra from the Mapping Nearby Galaxies at Apache Point Observatory (MaNGA) experiment.

 The gravitational redshift as an observable is a prominent and direct probe of gravity at cosmological scales \citep{Wojtak:2011ia, Alam2017}. We expect to see the effect of dark energy perturbations directly in the gravitational redshift signal since {$\Psi \propto \mu(k,z) \eta(k,z) \Delta_m$, see Eqs.\ (\ref{eq:Hamiltonian}) and (\ref{eq:slip}). Hence,} 
 \begin{multline}
 {C_\ell^{\delta z_\text{grav}} = 36 \pi^2 H_0^4 \Omega_m^2 \left[1 + \bar{z}(\chi_s)\right]^4} \\ {\times\int_0^\infty \! \frac{dk}{k^2} j_\ell^2(k\chi_s) \mu^2(k,\chi_s) \eta^2(k,\chi_s) P_{\Delta_m}(k,\chi_s)\,.}
 \end{multline}
 In Sec.\ \ref{sec:numerics} we study the effect of $k$-essence dark energy on the gravitational redshift signal.
\subsection{Shapiro time delay }\label{sec:shapiro}
In addition to 
{all the} effects 
discussed in the previous sections, the gravitational potential of large-scale structure 
{perturbs the interval of cosmic time while a photon traverses a given coordinate distance.}
This effect is known as Shapiro time delay {and} is first discussed and introduced 
in \citet{PhysRevLett.13.789} as a fourth test of general relativity. 

While the gravitational lensing is due to the gradient of the projected potentials along the {photon} trajectories {and is weighted by the lensing kernel}, the Shapiro time delay is {proportional} to the projected potential itself. As a result we expect {much} more signal at high {multipoles} from {the} lensing {convergence} compared to {the} Shapiro time delay, {due to the additional factor of $\sim \ell^2$.}

Due to Shapiro time delay the last scattering surface is a deformed sphere as different light rays travel through different gravitational potentials to reach us. All the photons {we receive from} the last scattering surface were released almost at the same time, so the time delay means that the photons {reaching us from different directions} have started at different distances from us. The modulation of the spherical surface due to Shapiro time delay is $\sim 1$ Mpc \citep{Hu:2001yq}.
The effect of Shapiro time delay on the {CMB} temperature and polarisation anisotropies is studied in \cite{Hu:2001yq}. They show that, while it is difficult to extract the Shapiro time delay signal from the data, {neglecting it} would introduce a systematic error. They argue that the Shapiro time delay should be considered {in order} to reduce the systematics in the analysis, especially for future high precision experiments. 

In \cite{Li:2019qkp} an estimator quadratic in the temperature and polarization fields is introduced to provide a map of the Shapiro time delays as a function of position on the sky. They show that the signal to noise ratio of this map could exceed unity for the dipole, so the signal could be used to provide an understanding of the Universe on the largest observable scales.

As discussed in \citet{Nusser2016}, {for tests of the equivalence principle at} high redshift {that rely on the Shapiro time delay effect}, potential fluctuations from the large scale structure of the Universe are at least two orders of magnitude larger than the gravitational potential of the Milky Way. {This suggests that the effect of dark energy on these potentials needs to be considered in order to model the Shapiro time delay accurately. From the last term in Eq.~(\ref{eq:position}) we get} 

\be
{\Delta\tau} = {\left(1 + \bar{z}\right) \Delta t =} -\int_{0}^{\chi_{s}}\!(\Phi+\Psi) \mathrm{d} \chi\,, \label{eq:Shapiro}
\ee
{and hence}
\begin{multline}
{C_\ell^{\Delta\tau} = 16 \pi^2 \int_0^\infty \! k^2 dk \int_0^{\chi_s} \! d\chi \int_0^{\chi_s} \! d\chi' j_\ell(k\chi) j_\ell(k\chi')} \\
{\times \left[1+\eta(k,\chi)\right] \left[1+\eta(k,\chi')\right] P_\Phi (k,\chi,\chi')\,.}
\end{multline}
{The dependence on $\mu(k,z)$ is evident from Eq.~(\ref{eq:Ppot}). This expression can be compared to Eq.~(\ref{eq:Clkappa}) where dark energy enters identically in the integrand, but the effect is weighted differently along the line of sight due to the lensing kernels. In our numerical analysis we obtain sky maps of the Shapiro time delay by directly solving Eq.~(\ref{eq:Shapiro}).}

\section{simulations} \label{sec:Nbody}
The results presented in this paper are based on {simulations performed with the relativistic $N$-body codes} \kev~\citep{Hassani:2019lmy}, {which includes non-linearly clustering dark energy in the form of a $k$-essence scalar field}, and \gev\ \citep{Adamek:2016zes} for $\Lambda$CDM and cases where linear dark energy perturbations {are sufficient} [see Fig.~1 in \citet{Hassani:2019lmy}]. We also compare our results from these two $N$-body codes with the linear Boltzmann code \class~\citep{Blas2011}.

In  \gev\ and \kev, for a fixed observer, {a ``thick'' approximate past light cone is constructed that encompasses a region sufficiently large to permit the reconstruction of the {true} light cone following the deformed photon trajectories {in post-processing} \citep{Adamek:2018rru}. The metric information for this thick light cone is saved} in spherical coordinates pixelised using {HEALP}ix \citep{Gorski2005}.
This lets us compute light-cone observables including weak gravitational lensing, ISW-RS, Shapiro time delay and gravitational redshift {in} post-processing. {In the present work we use a fast pixel-based method to solve the integrals of Eqs.~(\ref{eq:lensingphi}), (\ref{eq:isw}) and (\ref{eq:Shapiro}) in the Born approximation, given that post-Born corrections typically affect the angular power spectra only at very high multipoles. A comparison of this approach with ``exact'' ray tracing of the deformed geodesics is presented in \citet{Lepori:2020ifz} for the case of the weak lensing convergence [see also \citet{Pratten:2016dsm}] and justifies the use of the Born approximation for the multipoles discussed here. Note, however, that post-Born corrections can play an important role in higher-order correlation functions like the angular bispectrum.}

{For illustration, i}n Fig.~\ref{Fig:maps} the maps of each physical effect from a $\Lambda$CDM simulation are shown in the left panels. In the right panels the difference between the maps from $k$-essence with $w_0=-0.9$ and $c_s^2=10^{-7}$  and $k$-essence with $w_0=-0.9$ and $c_s^2=1$ are shown. 
 Since the seed number of the simulations are {identical}, {this shows} the effect of $k$-essence clustering on each map, because $k$-essence does not cluster {significantly} for the case with high speed of sound squared, $c_s^2=1$.
\begin{figure*}
\begin{minipage}{\textwidth}
\begin{tabular}{cc}
 \includegraphics[width=0.48\textwidth]{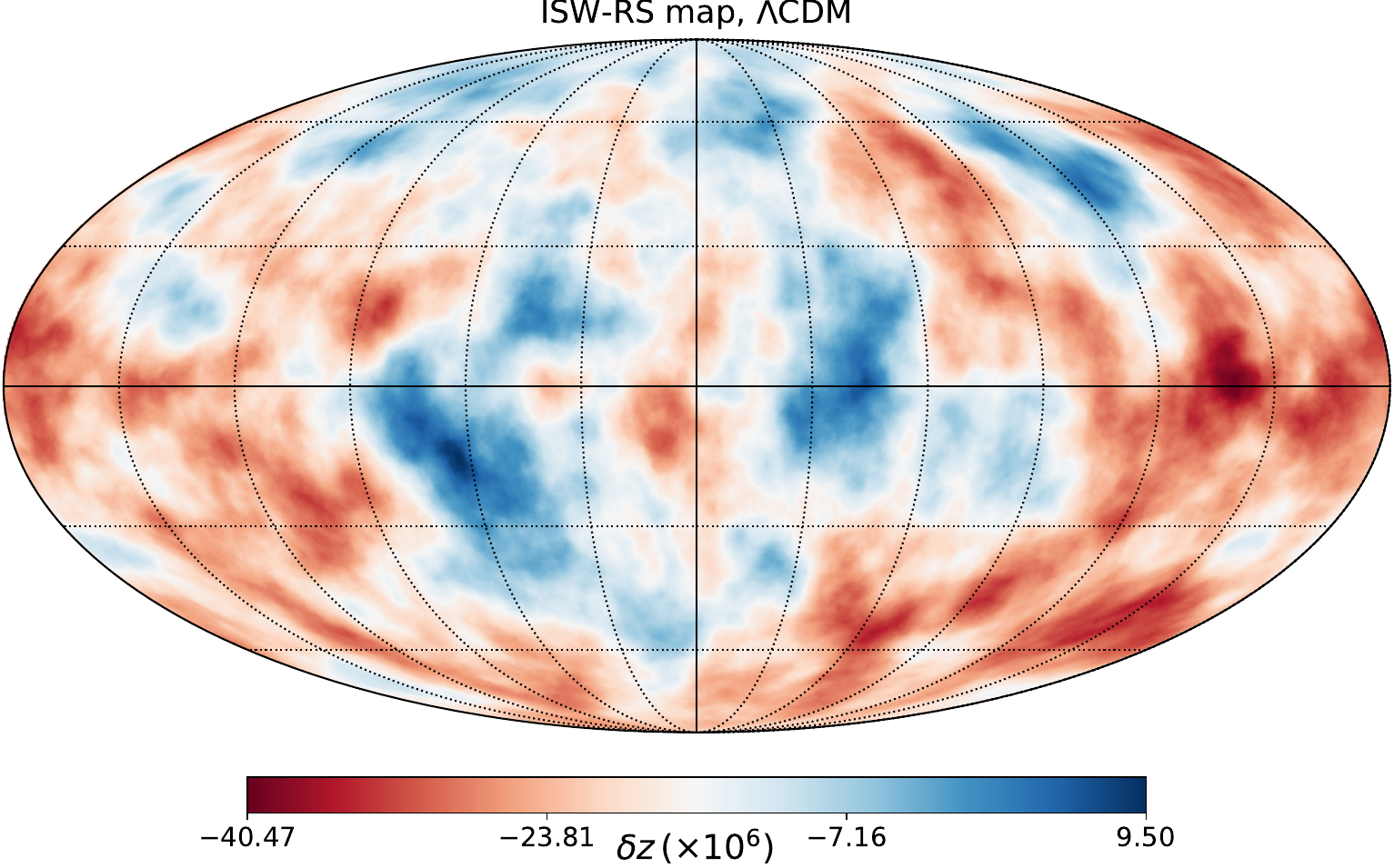} & \includegraphics[width=0.48\textwidth]{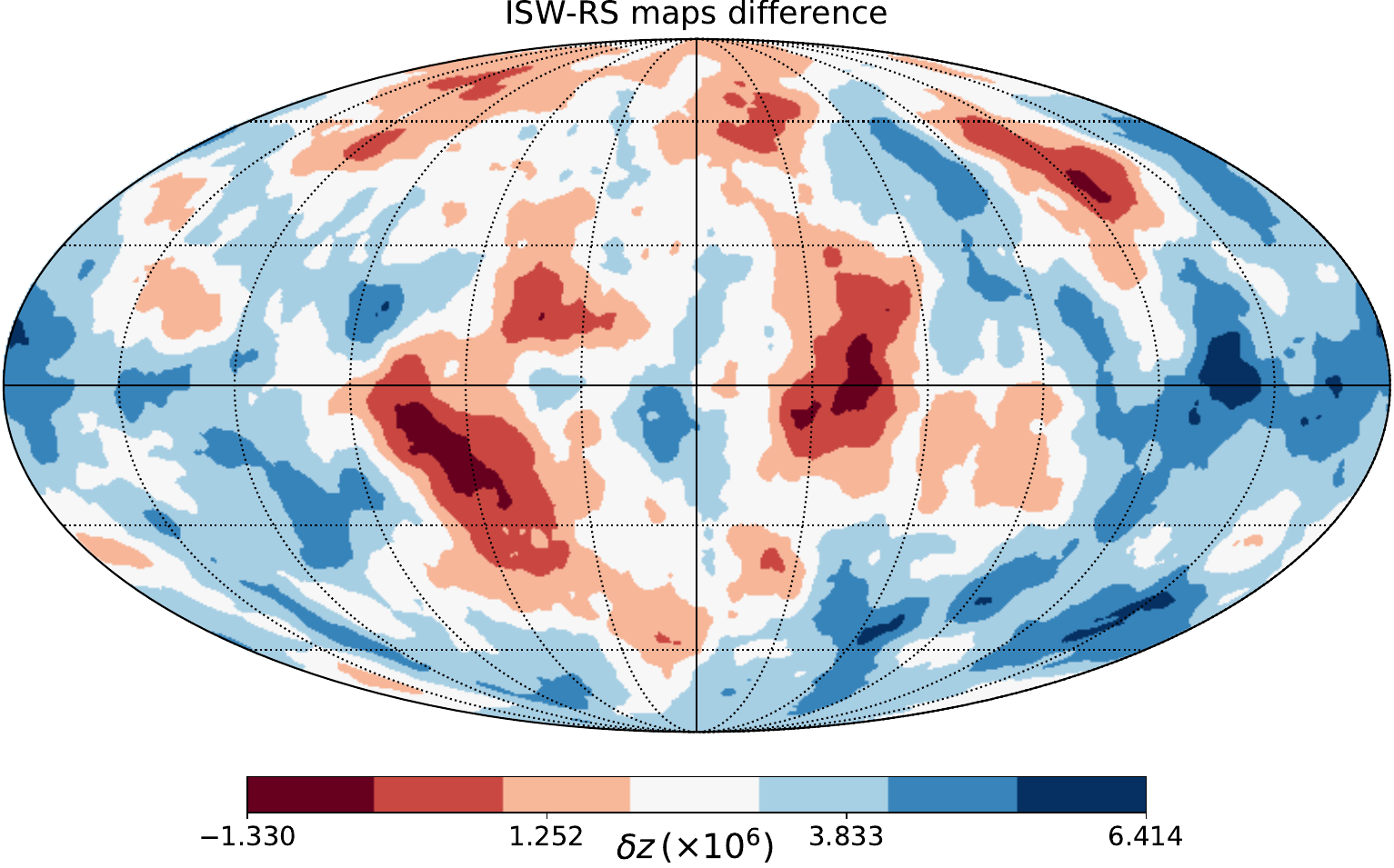} \\
 \\
  \includegraphics[width=0.48\textwidth]{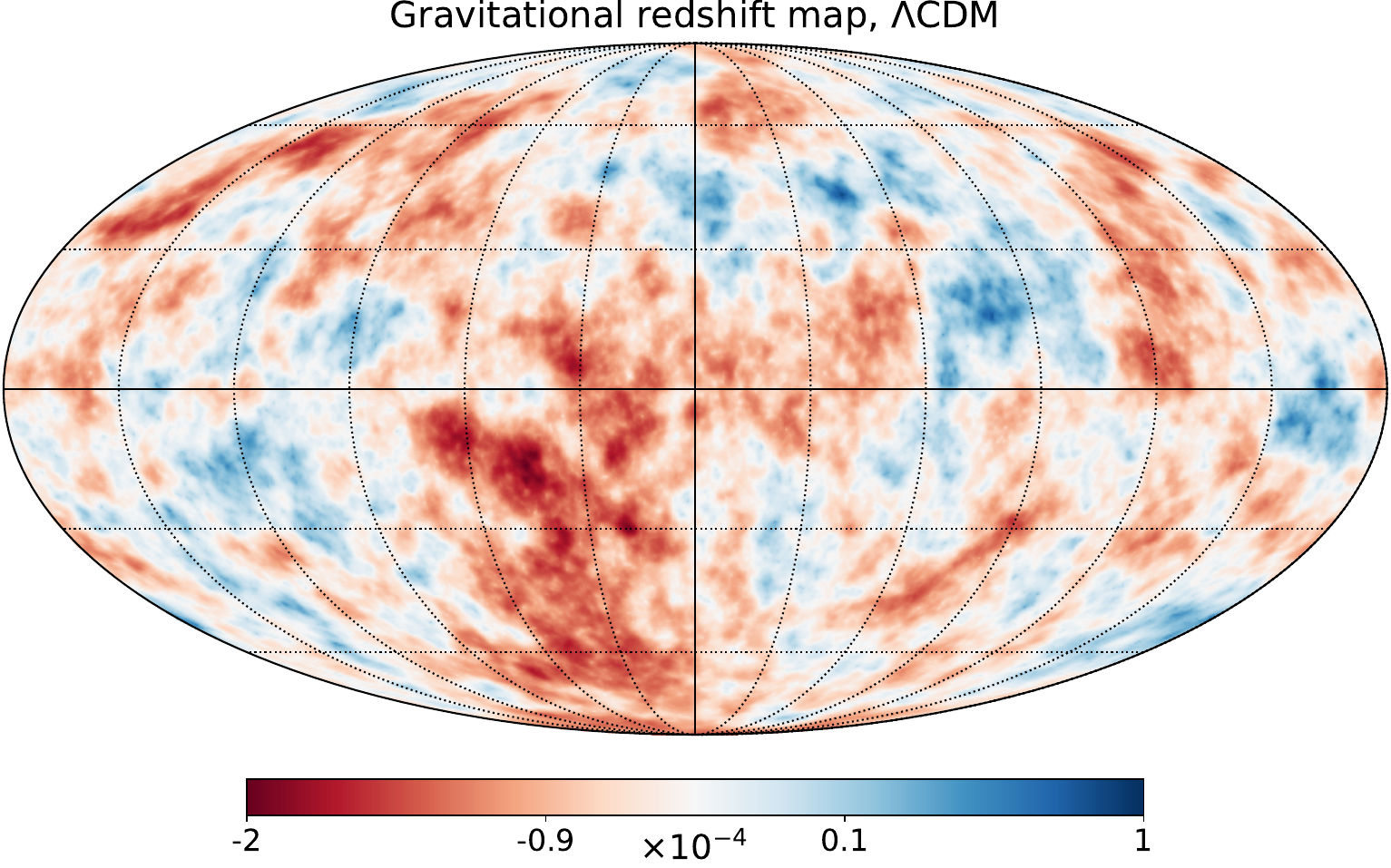} & \includegraphics[width=0.48\textwidth]{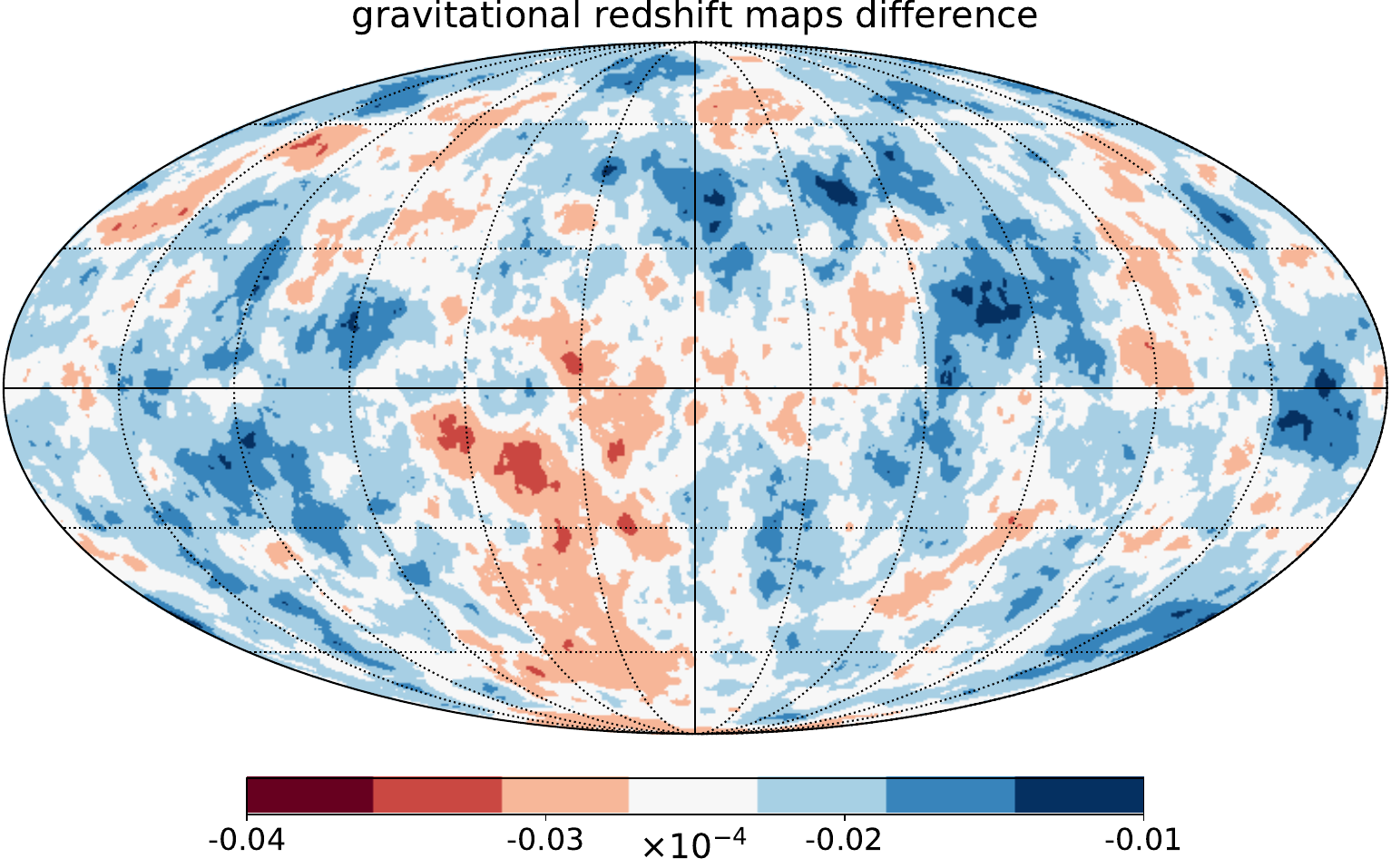} \\
  \\
 \includegraphics[width=0.48\textwidth]{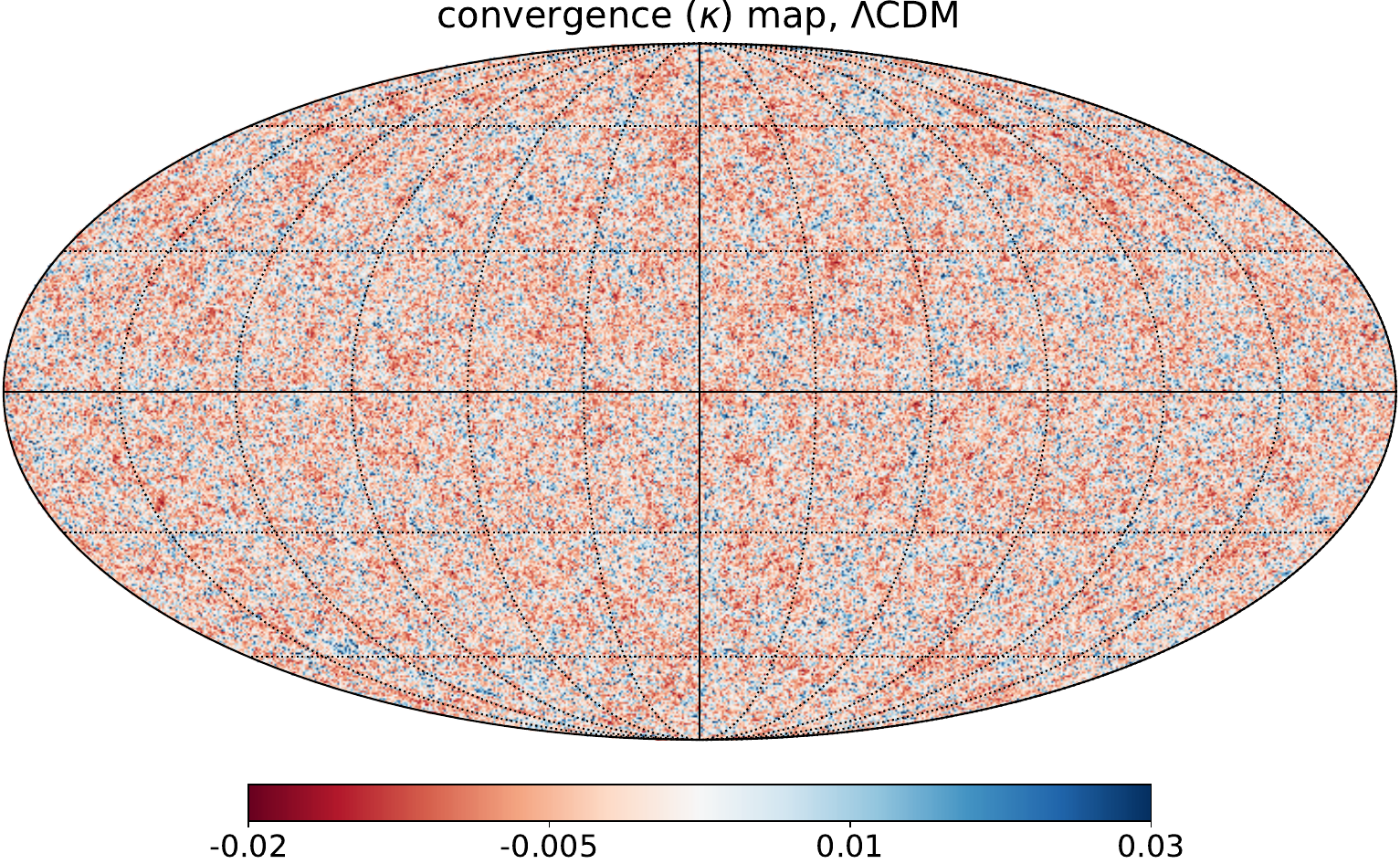} & \includegraphics[width=0.48\textwidth]{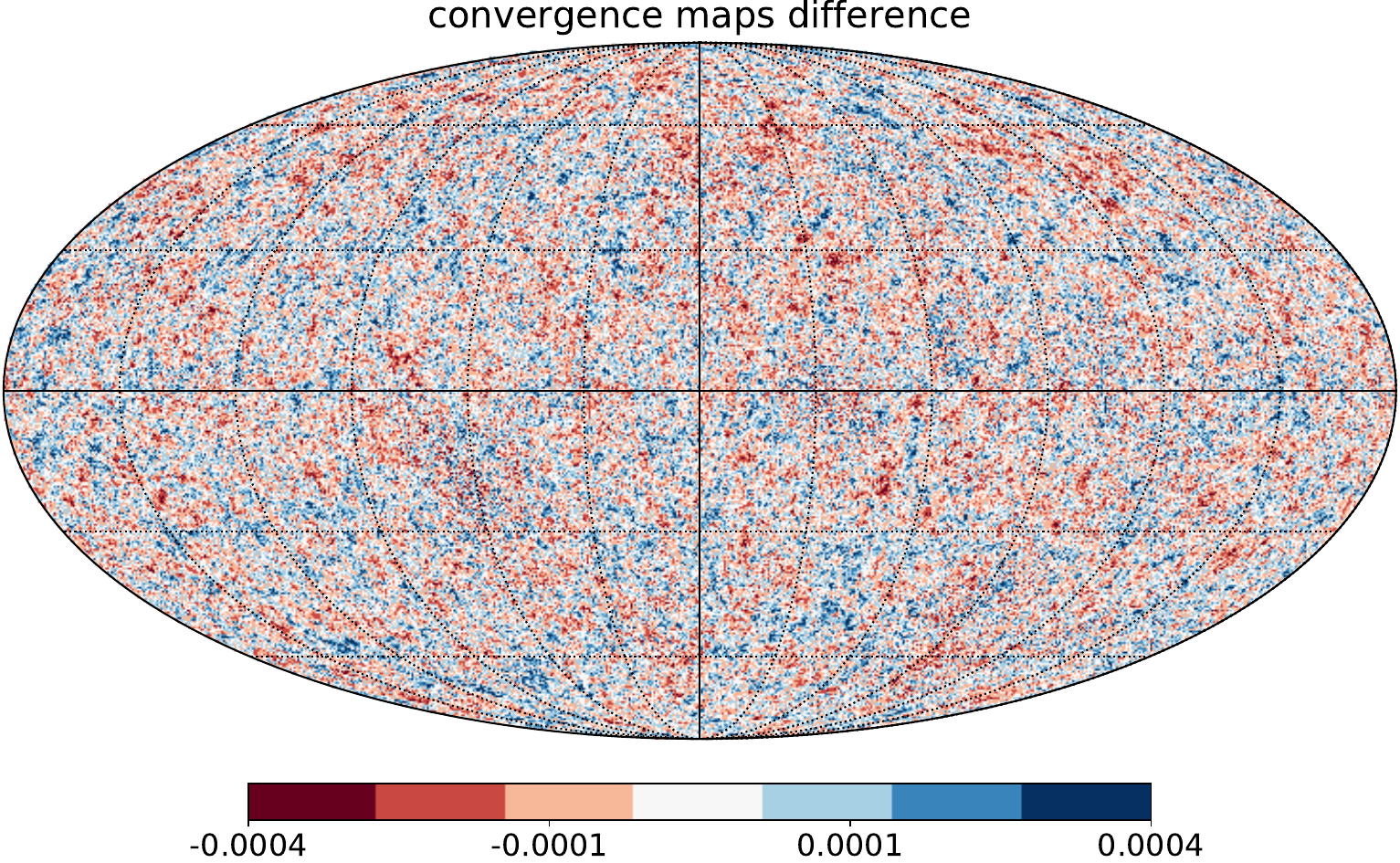}  \\
 \\
  \includegraphics[width=0.48\textwidth]{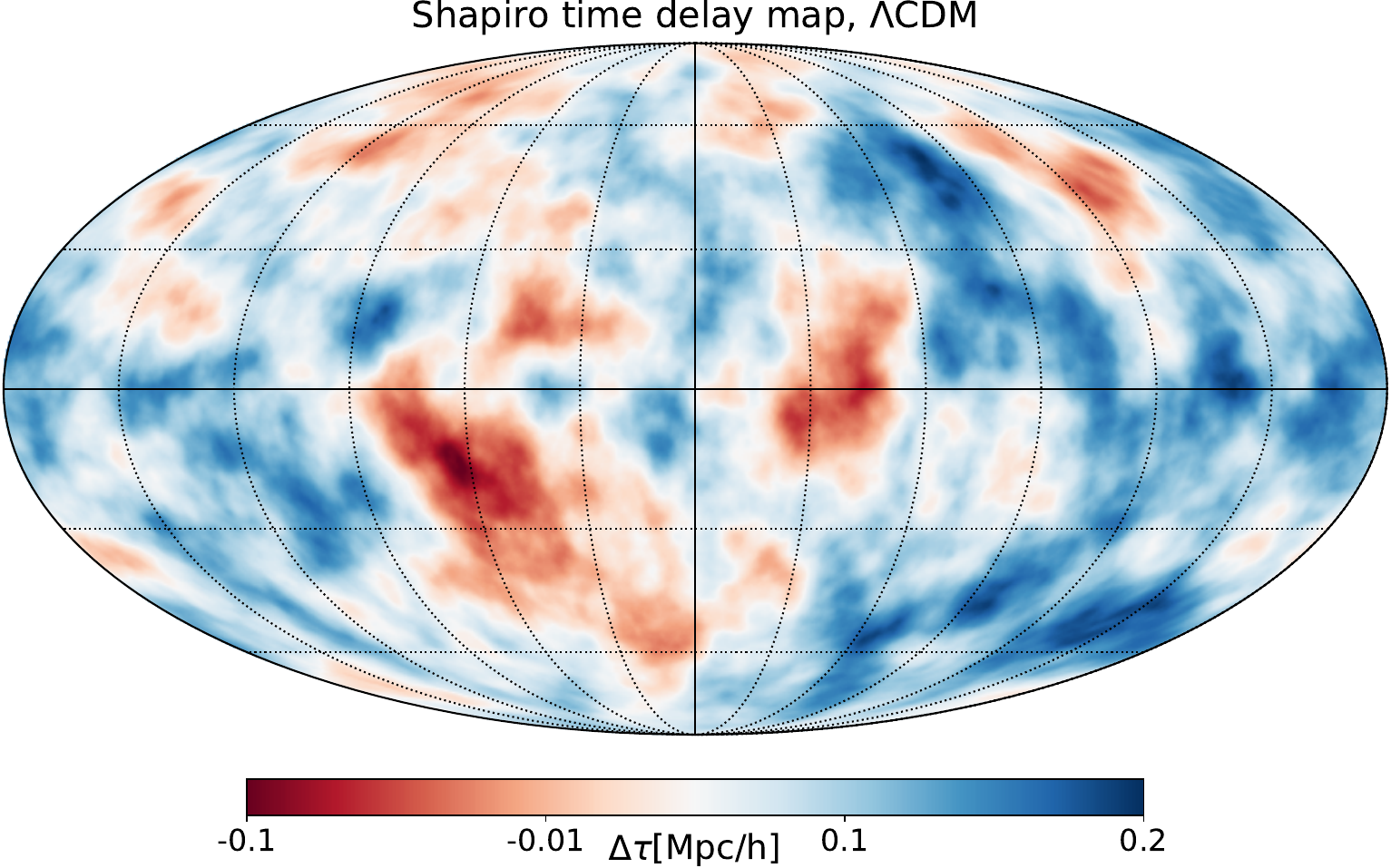} & \includegraphics[width=0.48\textwidth]{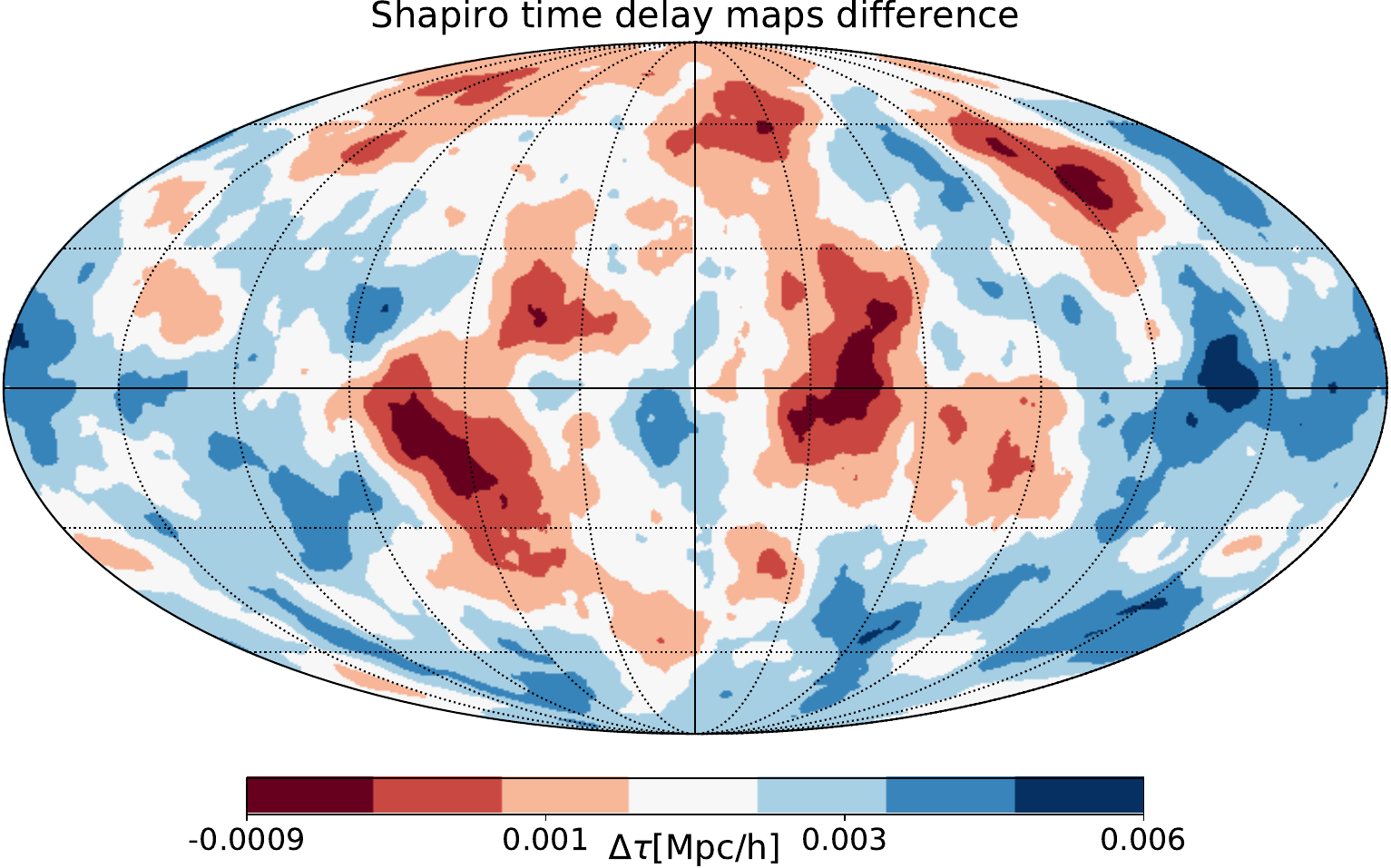}
\end{tabular}
\end{minipage}
\caption{\label{fig:maps} \small {\bf \emph{Left:} }Full-sky maps of the ISW-RS temperature anisotropy, gravitational redshift, convergence and Shapiro time delay from a $\Lambda$CDM simulation using \gev, integrat{ed} to $z\sim 0.8$ {are} shown. {\bf \emph{Right:}} Difference between the maps for $k$-essence {with} $c_s^2=10^{-7}$ {and} $k$-essence {with} $c_s^2=1${, both simulated with} \kev. {All simulations used the same seed to generate the random initial conditions and hence the difference maps indicate} the importance of $k$-essence clustering.
}
\label{Fig:maps}
\end{figure*}

In our simulations we place the observer in the corner of the box at position (0,0,0) in Cartesian coordinates. 
We store data on the past light cone of the observer {on the full sky out to a comoving distance of $\chi_s = 2015\, \text{Mpc}/h$, which corresponds} 
approximately to $z=0.8${, and out to a comoving distance of $\chi_s = 4690\, \text{Mpc}/h$ or a corresponding approximate redshift of} 
$z=3.3$ for {a pencil beam covering a sky area of $1932$ sq.\ deg.\ in the direction of the diagonal of the simulation box.}

The simulation {boxes have a comoving size of} $L = 4032\, \text{Mpc}/h$. {The fields evolve on a grid with} $N_{\text {grid}} = 4608^3$ {grid points, and the matter phase space is sampled by} $N_{\text{pcl}} = 4608^3$ (i.e.\ about 100 billion) {particles}. {We use the following cosmological parameters for all runs:}
the amplitude of  scalar perturbations {is set to} $A_s = 2.1 \times 10^{-9}$ at the pivot {scale} $k_p = 0.05$ Mpc$^{-1}$, the scalar spectral index is $n_s = 0.96$, the Hubble parameter $h = 0.67$, cold dark matter and baryon densities are, respectively, $\omega_{\text{cdm}} = \Omega_{\text{cdm}} h^2 =0.121203 $ and $\omega_{\text{b}} = \Omega_{\text{b}} h^2 
 =0.021996 $, and the CMB temperature $T_{\text{CMB}} = 2.7255$ K. We also {include} two massive neutrino species with masses $m_1 =  0.00868907$ eV and $m_2 = 0.05$ eV with temperature parameter $T_{\text{C}\nu\text{B}}  = 1.95176$ K, as well as $N_{ur} = 1.0196$ massless neutrinos \citep{Adamek:2017uiq}. The effect of {neutrino and} radiation perturbations is approximated using the linear transfer function{s} from \class\ {as described in \citet{Brandbyge:2008js} and \citet{Adamek:2017grt}, respectively}. {We only consider spatially flat universes, $\Omega_k = 0$, so that the dark energy density parameter is given by} $\Omega_{\text{DE}} = 1 - \sum_X \Omega_X$ where the sum goes over all species except dark energy. {For the simulations where the dark energy is not a cosmological constant, we use a constant equation of state relation given by} $w_0= -0.9$ and $w_a=0$.

 The initial conditions for the simulations are {set} using the linear transfer functions from \class\ at $z=100$ and all the simulations are run with the same seed number which helps us to compare the results.

{For our \kess\ cosmology we consider three different choices for the speed of sound: $c_s^2 = 1$, $c_s^2 = 10^{-4}$ and $c_s^2 = 10^{-7}$. Strong non-linear clustering of the dark energy field is only expected in the last case, and to study this specific aspect, we run two separate simulations for this case. In one simulation we use the code \gev\ which approximates the dark energy perturbations by their solution from linear theory, see \citet{Hassani:2019lmy} for more details. In the other simulation we use the code \kev\ in order to keep track of the dark energy field's response to the non-linear clustering of matter. For the higher values of the speed of sound we only run \kev. We also do a reference run for the $\Lambda$CDM model with \gev\ -- here there is no difference between the two codes because dark energy perturbations are absent.} 

\begin{figure}
    \includegraphics[width=\columnwidth]{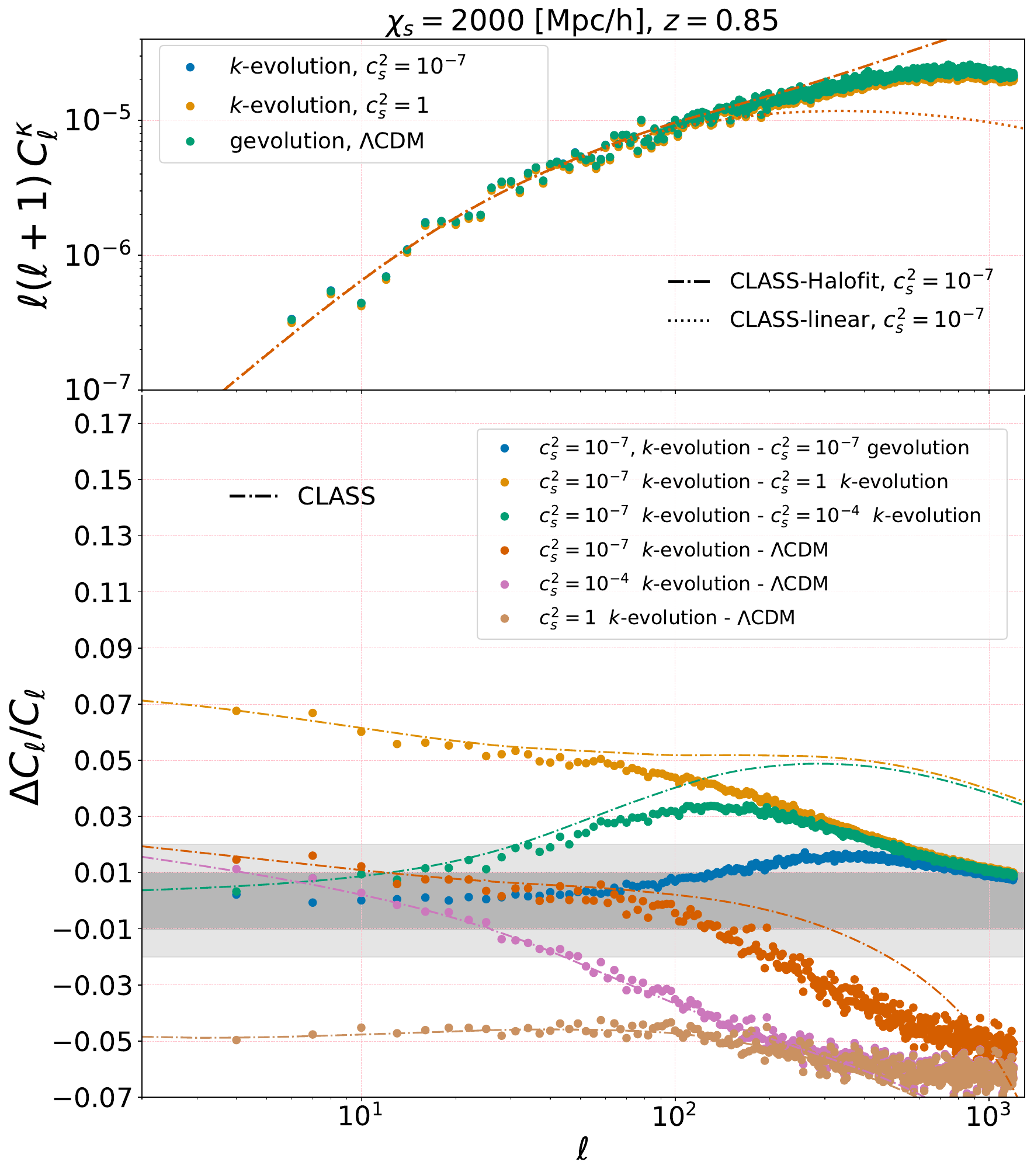}
       \caption{\textbf{\emph{Top panel:}} The angular power spectra of {the} lensing convergence $\kappa$ {at a source distance of $\chi_s = 2000$ Mpc$/h$ for some of our $N$-body simulations, as well as predictions from \class\ with and without the Halofit prescription.}
       {Based on the convergence test of App.\ \ref{app:convergence} we can trust the convergence spectrum here to $\ell\approx200$, which is roughly where the dots start to deviate from the CLASS-Halofit curve. Relative spectra remain accurate to much higher $\ell$.}
 \textbf{\emph{Bottom panel:}} {Data points show t}he relative difference between the convergence power spectra from {our} $N$-body simulations {for different models, whereas dashed} 
 lines represent the {corresponding predictions from \class\ with Halofit, shown in matching colours.} 
 The {blue points} 
 show the impact of \kess\ non-linearities by comparing non-linear \kess\ from \kev\ with {its} linear implementation in \gev.}
 \label{fig:kappa}
\end{figure}
\begin{figure}
    \includegraphics[width=\columnwidth]{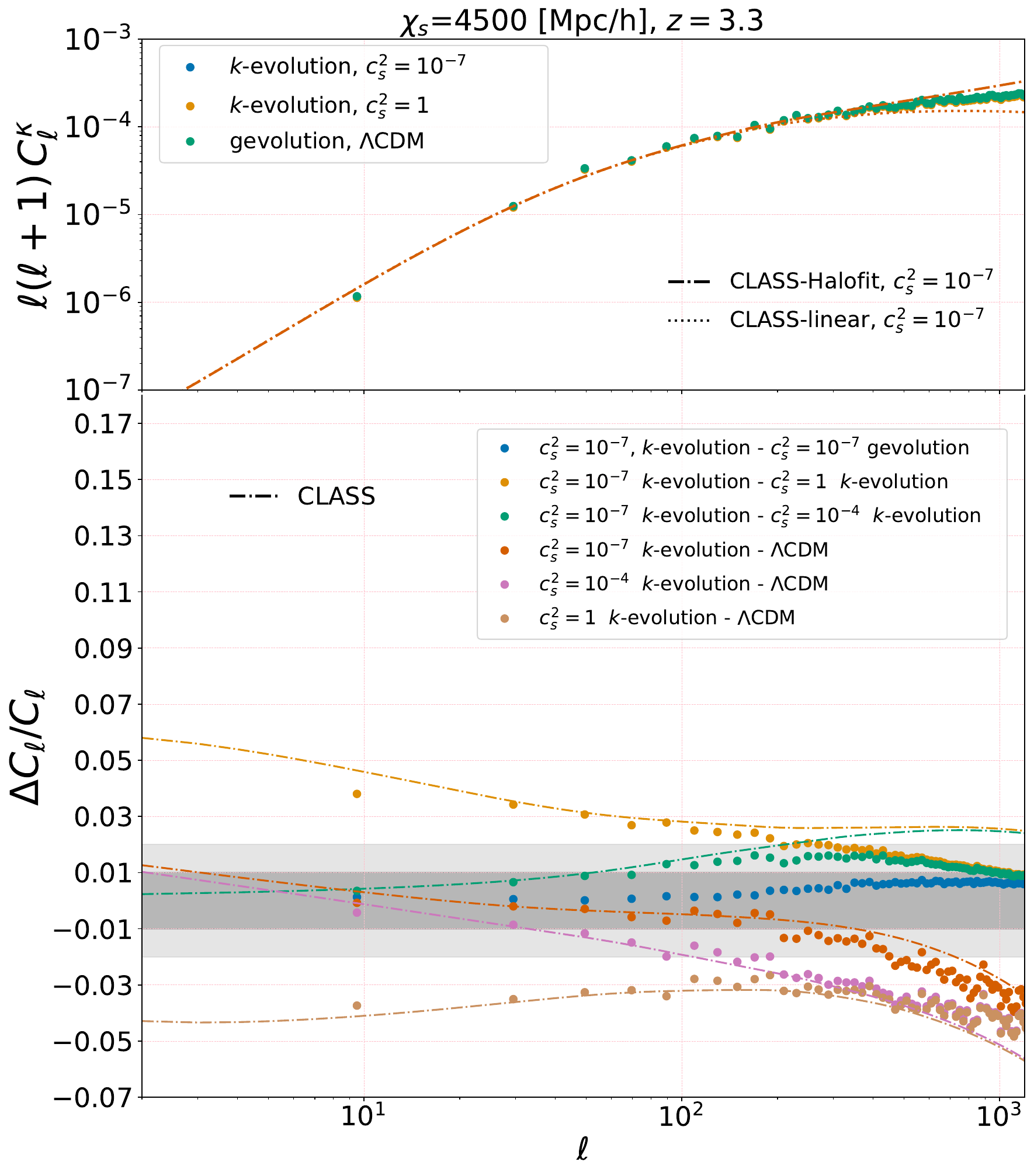}
       \caption{
       {Same as Fig.~\ref{fig:kappa}, but for a source distance of $\chi_s = 4500$ Mpc$/h$, corresponding roughly to $z \approx 3.3$, and using $N$-body simulation data that covers $\sim 5\%$ of the sky.} {Again based on App.\ \ref{app:convergence} we can trust the dots in the top panel to $\ell\approx 500$, while the relative spectra shown in the bottom panel are again valid to higher $\ell$.}}
        \label{fig:kappa_4000}
\end{figure}

\section{Numerical results}
\label{sec:numerics}
In this section we show the results of {our} numerical simulations. 
{Once we have generated the maps with our pixel-based numerical integrator we use \textsc{anafast} from the \textsc{python} package of HEALPix to compute}
the angular power spectra and cross power spectra. {However,} for the pencil beam maps {that cover only about 5\% of the sky we instead}
use the pseudo-$C_{\ell}$ estimator {of} \citet{Wandelt2001, Szapudi:2000xj} to obtain the power spectra in an unbiased way. To do so we use {the} \textsc{PolSpice} package to compute the angular power spectra and cross power spectra of the masked maps.

\begin{figure}
    \includegraphics[width=\columnwidth]{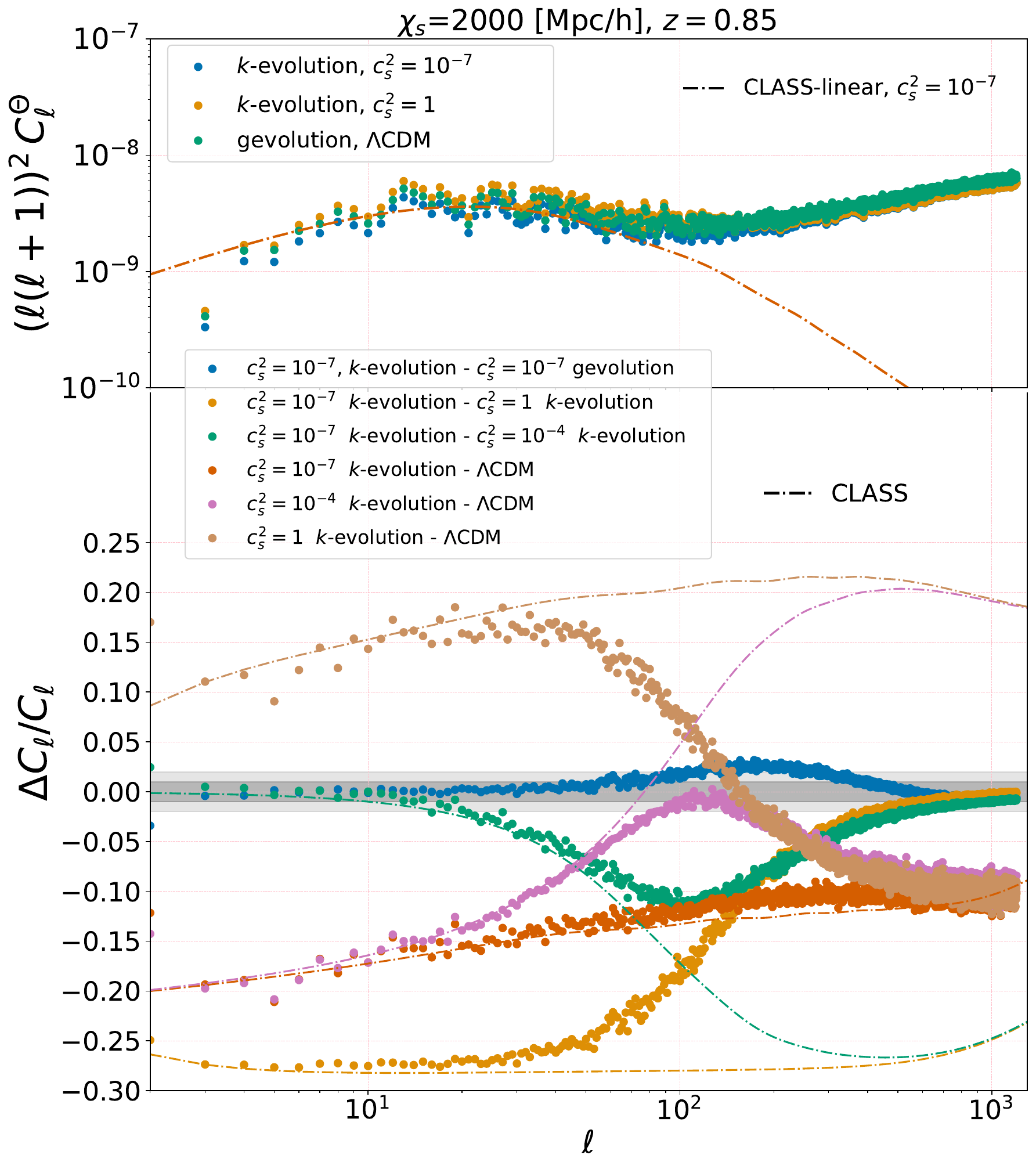}
      \caption{\textbf{\emph{Top panel:}} The late-ISW-RS angular power spectra for different cosmologies integrating to ${\chi_s} = 2000$ Mpc$/h$ are shown. Note that \class\ {only predicts the linear ISW signal without any non-linear correction applied.} {The convergence study of App. \ref{app:convergence} indicates that these spectra are reliable to $\ell\approx 500$, the relative spectra are again valid to higher $\ell$.} 
      \textbf{\emph{Bottom panel:}} The relative difference between different models in the ISW-RS power spectra are shown. At large scales the result from $N$-body codes agree with \class, however, {already at $\ell < 100$} 
      the linear and non-linear curves diverge due to the non-linear RS effect.}
           \label{fig:ISW_2000}
\end{figure}
\begin{figure}
    \includegraphics[width=\columnwidth]{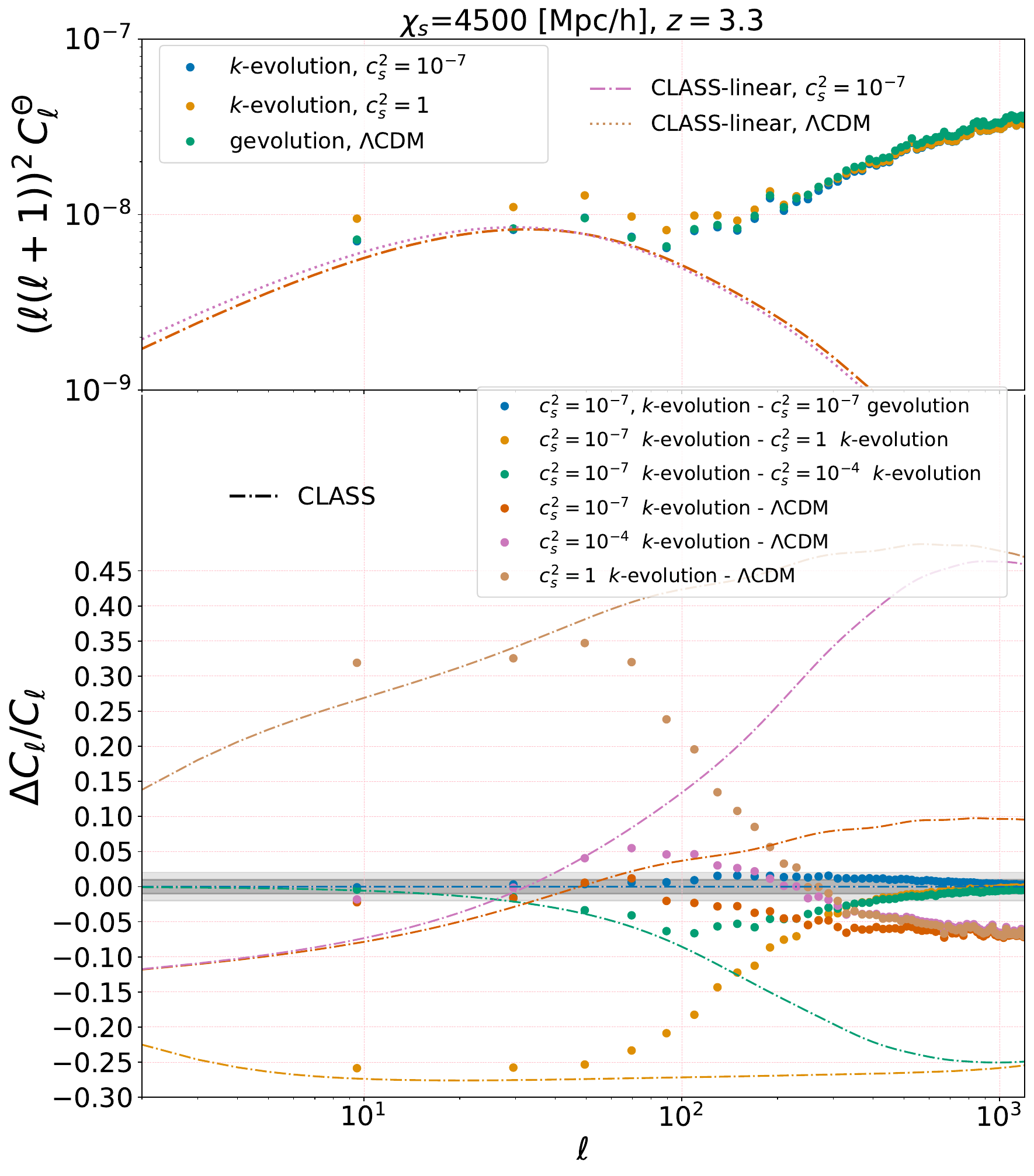}
      \caption{
      Same as Fig.~\ref{fig:ISW_2000}, but for a source distance of $\chi_s = 4500$ Mpc$/h$, corresponding roughly to $z \approx 3.3$, and using $N$-body simulation data that covers $\sim 5\%$ of the sky. {From the convergence study in the appendix we conclude that these spectra are reliable over the full range of scales shown.}}
      \label{fig:ISW_4000}
\end{figure}

\subsection{Weak gravitational lensing}
In {the top panel of} Fig.~\ref{fig:kappa} the angular power spectra for {the} lensing convergence from \kev, \gev~and \class~integrating up to ${\chi_s}=2000$ Mpc$/h$ are shown{, which corresponds to a source redshift of approximately $z = 0.85$}. On the top {the data points} show the results for $\ell (\ell +1)C_{\ell}^{\kappa}$ {of} two simulations {with} $c_s^2 = 10^{-7}$ and $c_s^2 =1$, 
{as well as for} $\Lambda$CDM. 
{In addition,} results for \class\ 
are shown 
{for the case $c_s^2 = 10^{-7}$ as} dash-dotted and dotted lines{, respectively using linear theory alone or Halofit \citep{Takahashi:2012em} to model the non-linear matter power spectrum}.

The effect of matter non-linearities appears in the convergence power spectrum at $\ell \sim 100$ {where} both {Halofit} and $N$-body simulation data start to diverge from {the} linear {prediction}. In the bottom {panel} {of Fig.~\ref{fig:kappa}} the relative difference between the convergence power spectra {of different models} are shown {(we always keep $\chi_s$ fixed which means that the source redshift can vary by a small amount depending on the background cosmology)}. This {quantifies} the effect of dark energy at different levels and {can be compared} to the prediction from \class. The {data points show again the numerical results from our} $N$-body {simulations, whereas} 
the \class/{Halofit} results {for the same model comparisons are plotted as} 
dashed lines {using the same colours}. 
{Note that the comparison between \kev\ and \gev\ for the case $c_s^2 = 10^{-7}$ has no corresponding prediction from \class, and quantifies the importance of the nonlinear modelling of \kess\ within the $N$-body simulations. To be specific, our convention is $\Delta C_\ell / C_\ell = (C_\ell^\text{(dataset 1)} - C_\ell^\text{(dataset 2)}) / C_\ell^\text{(dataset 2)}$, where ``dataset 1'' and ``dataset 2'' are indicated in the legend of each figure.}

At large scales, as expected, the results from \class\ and $N$-body simulations agree very well, while at smaller scales some deviations are seen.
The deviation of {the} $N$-body results from {the} \class{/H}alofit prediction occurs mainly when there is dark energy clustering which is more pronounced for the cases with low speed of sound. {Indeed, comparing the case $c_s^2 = 1$ with $\Lambda$CDM we find that the \class/Halofit prediction is accurate at least up to $\ell \sim 1000$, whereas for the lowest speed of sound, $c_s^2 = 10^{-7}$, significant disagreement between \class/Halofit and $N$-body simulations appears already for multipoles $\ell \gtrsim 100$.}
This is in complete agreement with our results in
\citet{Hassani:2019wed} where we show that \class/Halofit gives the right clustering function $\mu(k,z)$ for high speed of sound.

{Our comparison between the two simulation methods (\kev\ and \gev) for the case $c_s^2 = 10^{-7}$ shows that the non-linear response of \kess\ to matter clustering produces an effect of $\sim 2\%$ in the convergence power spectrum at multipoles in the range of a few hundred.}

According to Fig.~\ref{fig:kappa} the effect of dark energy
{clustering can cancel the effect of the}
background evolution {with} $w_0 = -0.9$ {on the weak-lensing signal at large angular scales: the $\sim 5\%$ suppression when going from $\Lambda$CDM to $w$CDM with $c_s^2 = 1$ is compensated by an amplification of similar size when going from $c_s^2 = 1$ to low speed of sound. The scale up to which this cancellation works is set by the sound horizon, and at very non-linear scales there is always some residual suppression.} {In our scenario} {the cancellation happens only for the case $w_0>-1$ when 
the linear growth rate is suppressed ({see} Fig~\ref{fig:BG}), while for the case with $w_0<-1$, where the growth rate is enhanced compared to $\Lambda$CDM, the background evolution and the dark energy clustering effects {would} act in a similar way.}

Fig.~\ref{fig:kappa_4000} is similar to Fig.~\ref{fig:kappa} 
but {shows results} from the pencil beam map integrating to {a} higher {source distance of $\chi_s = 4500$ Mpc$/h$, corresponding approximately to} redshift $z=3.3$. As we have access to only  $\sim 5\%$ of the sky {in this case}, we have no information {about the} angular power spectra at low $\ell$. 
The \class/Halofit and $N$-body simulation data agree better compared to the lower redshift results, as the scale {where the finite resolution of the simulation affects the result is shifted to higher multipoles, see Appendix \ref{app:convergence} [cf.\ also Appendix C of \citet{Lepori:2020ifz}].} 
Moreover, the relative difference between the angular power spectra is 
{less substantial}
compared to lower redshift result which comes from the fact that dark energy starts to dominate at lower redshift and integrating to higher redshifts {therefore effectively} 
dilutes the signal.  

{The overall signal amplitude for} lensing is {larger} at higher redshift owing to the fact that it {is} an integrated effect. Thus, although the {relative} effect of dark energy clustering {is lower} 
at higher redshifts, 
the detectability of the signal is larger, and combining high and low redshift lensing {data still significantly} increases the signal to noise ratio.

\begin{figure}
    \includegraphics[width=\columnwidth]{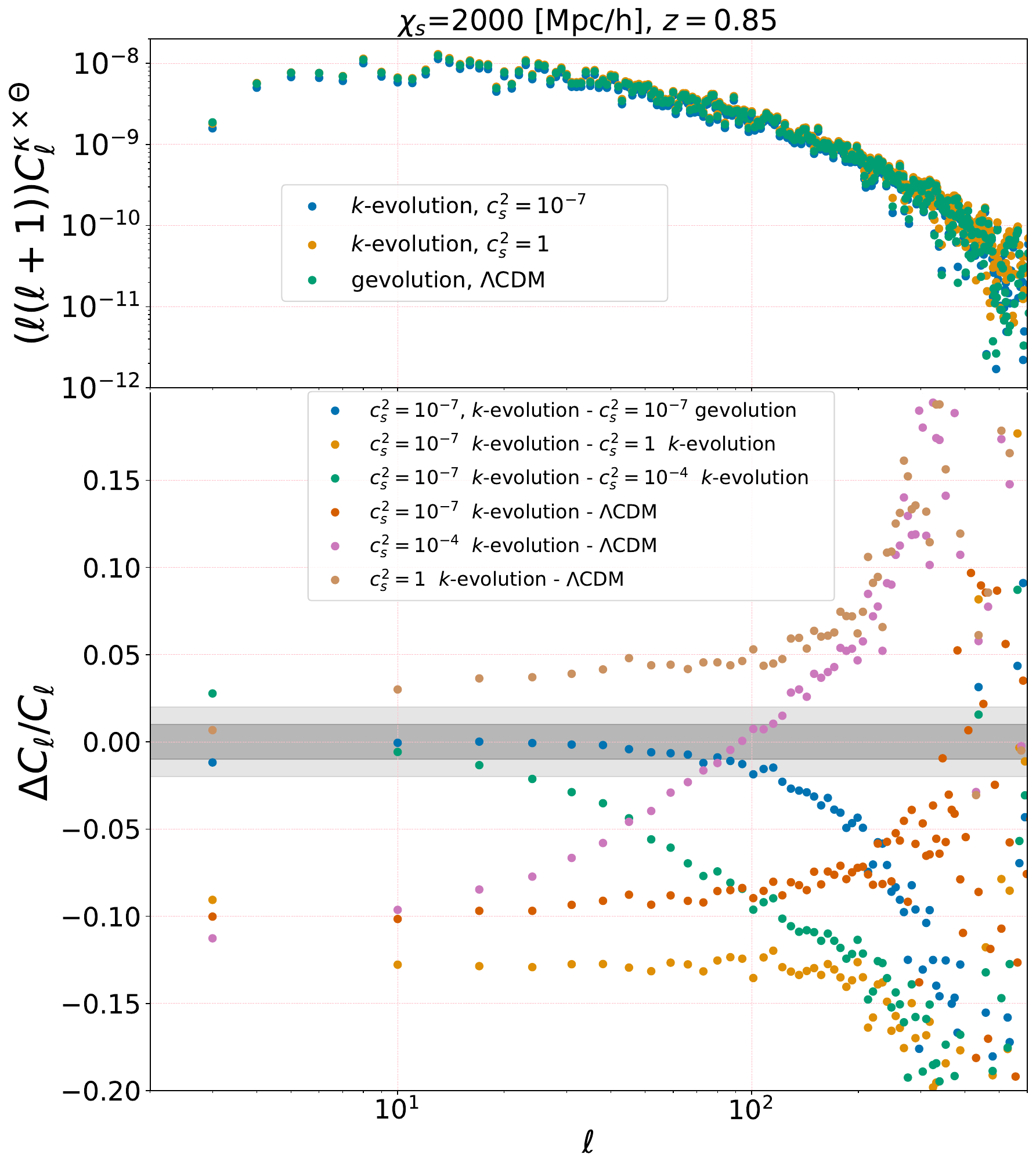}
       \caption{\textbf{\emph{Top panel:}} The cross power spectra of lensing convergence $\kappa$ and ISW-RS integrated to comoving distance ${\chi_s}=2000$ Mpc$/h$ from \kev\ with two speeds of sound $c_s^2 = 10^{-7}$ and $c_s^2=1$ {are shown, together with our} $\Lambda$CDM {reference run}. \textbf{\emph{Bottom panel:}} The relative difference between {the cross} power spectra 
       {for several models is shown.}}
       \label{fig:ISW-kappa_2000}
\end{figure}
\begin{figure}
    \includegraphics[width=\columnwidth]{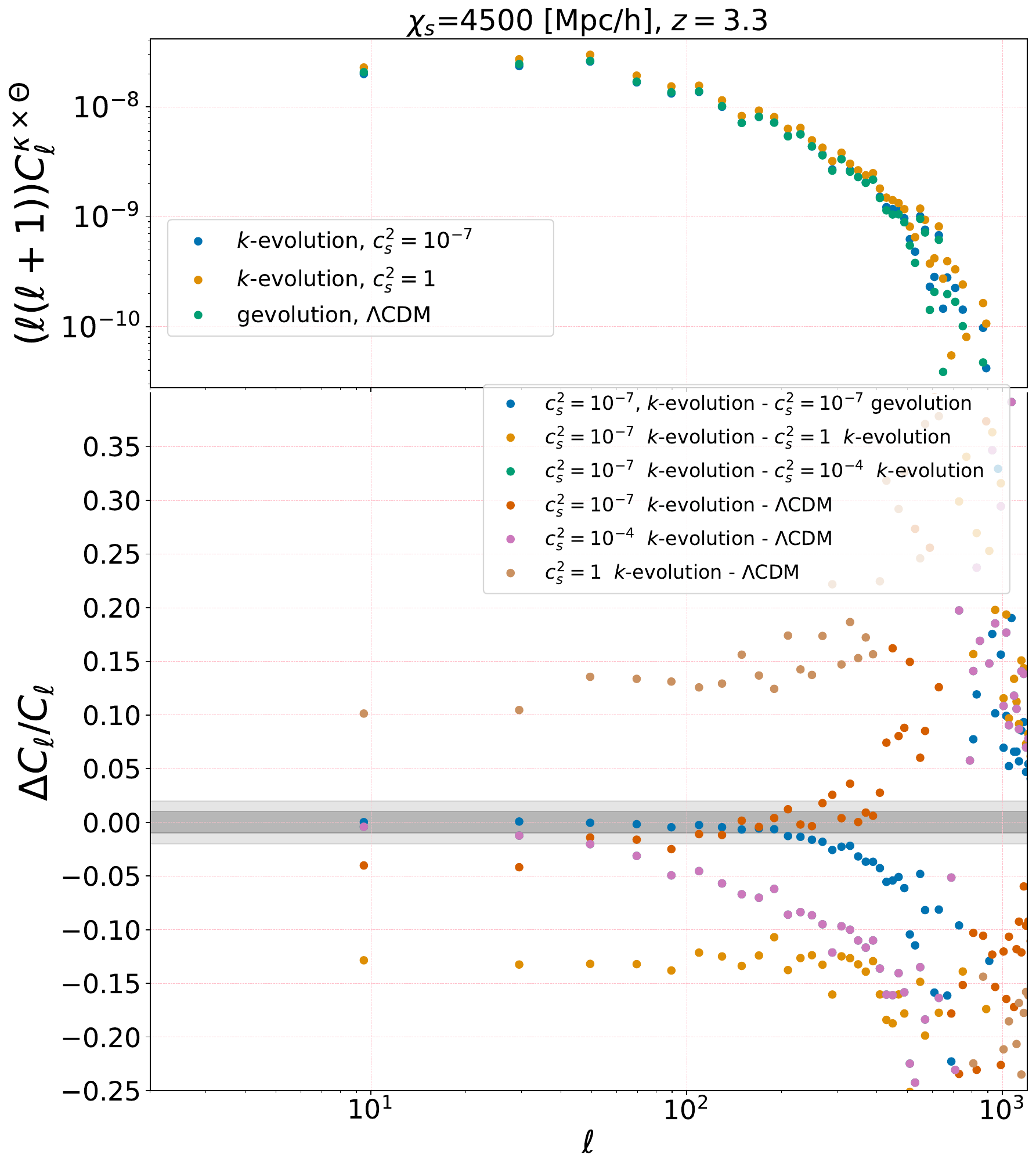}
       \caption{
       {Same as Fig.~\ref{fig:ISW-kappa_2000}, but for a source distance of $\chi_s = 4500$ Mpc$/h$, corresponding roughly to $z \approx 3.3$, and using $N$-body simulation data that covers $\sim 5\%$ of the sky.}}
      \label{fig:ISW-kappa_4000}
\end{figure}

\subsection{ISW-RS effect}
In Figs.~\ref{fig:ISW_2000} and \ref{fig:ISW_4000}, the ISW-RS angular power spectra from {our $N$-body simulations} and \class\ are shown and {the different dark energy models are} compared in the bottom {panel} of each figure. The ISW-RS signal is very sensitive to the clustering of dark energy and also its background evolution. Comparing dark energy with low and high speeds of sound with $\Lambda$CDM, one {finds a} huge impact $\sim35\%$ at $z=3.3$ and $\sim 30\%$ at $z=0.85$ from dark energy clustering which makes {the} ISW-RS signal an excellent probe of this model.

It {is also} interesting to see that the non-linear RS effect starts to have impact at larger scales than 
the scale of non-linearity in {the} lensing signal in Fig.~\ref{fig:kappa_4000} which is also observed and discussed in \citep{Cai2009, Adamek:2019vko}. To {verify} that {the} non-linear RS {indeed appears} 
in lower moments 
we design a {numerical} experiment to decrease the non-linear RS signal by decreasing $A_s$, the amplitude of scalar perturbations. {This shifts the non-linear scale and allows us to separate the linear ISW effect from higher-order corrections.} 
Our numerical results{, shown} in Appendix~\ref{app:NLRS}, verify that the non-linear RS effect is responsible for the deviation 
at lower $\ell$ compared to 
{lensing. This is maybe not too surprising given that the lensing kernel suppresses contributions from the vicinity of the observer which would be projected dominantly to low multipoles.}

The non-linear RS effect is not implemented in 
\class\ and as a result we a see power drop at ${\ell \sim 100}$ in the linear power spectra (top {panels}). Moreover{, it is} interesting to see that unlike the lensing power spectra, the effect of clustering dark energy and background evolution do not cancel each other in {the} ISW-RS power spectra and 
we obtain $\sim -20\%$ relative difference between $c_s^2=10^{-7}$ and $\Lambda$CDM. 

{As seen i}n Fig.~\ref{fig:ISW_4000} 
the signal itself is stronger at higher redshift as it is an integrated signal. 
{However,} depending on the parameters the relative difference can change positively or negatively compared to the lower redshift comparison. {This is because more linear ISW and less non-linear effect is accumulated at high redshift.}

As 
explained in Sec.~\ref{sec:ISW_RS} the direct detection of ISW-RS signal is difficult and {it} is {therefore} usually detected indirectly, e.g.\ via cross correlation with other quantities{. As an example}, we
report {the} convergence-ISW-RS cross power spectra {in} 
Fig.~\ref{fig:ISW-kappa_2000} and Fig.~\ref{fig:ISW-kappa_4000}. 
Fig.~\ref{fig:ISW-kappa_alm_2000} {shows the cross-correlation coefficient, i.e.\ the cross power normalised to the r.m.s.\ of each individual signal.} 
The {effects of} dark energy clustering and expansion history in {the} 
cross power spectra {follows similar trends as seen for the individual probes.} 
Interestingly, the cross-correlation coefficient saturates to $\sim 0.8$ at low $\ell$ almost independently of the dark energy model. This is because virtually all the effect is taken up by the normalisation. However, the non-linear evolution is different in different models, {which leads to an absolute change, $\Delta C_{\ell}$, in the cross-correlation coefficient of a few percentage points at high multipoles.} 

\begin{figure}
    \includegraphics[width=\columnwidth]{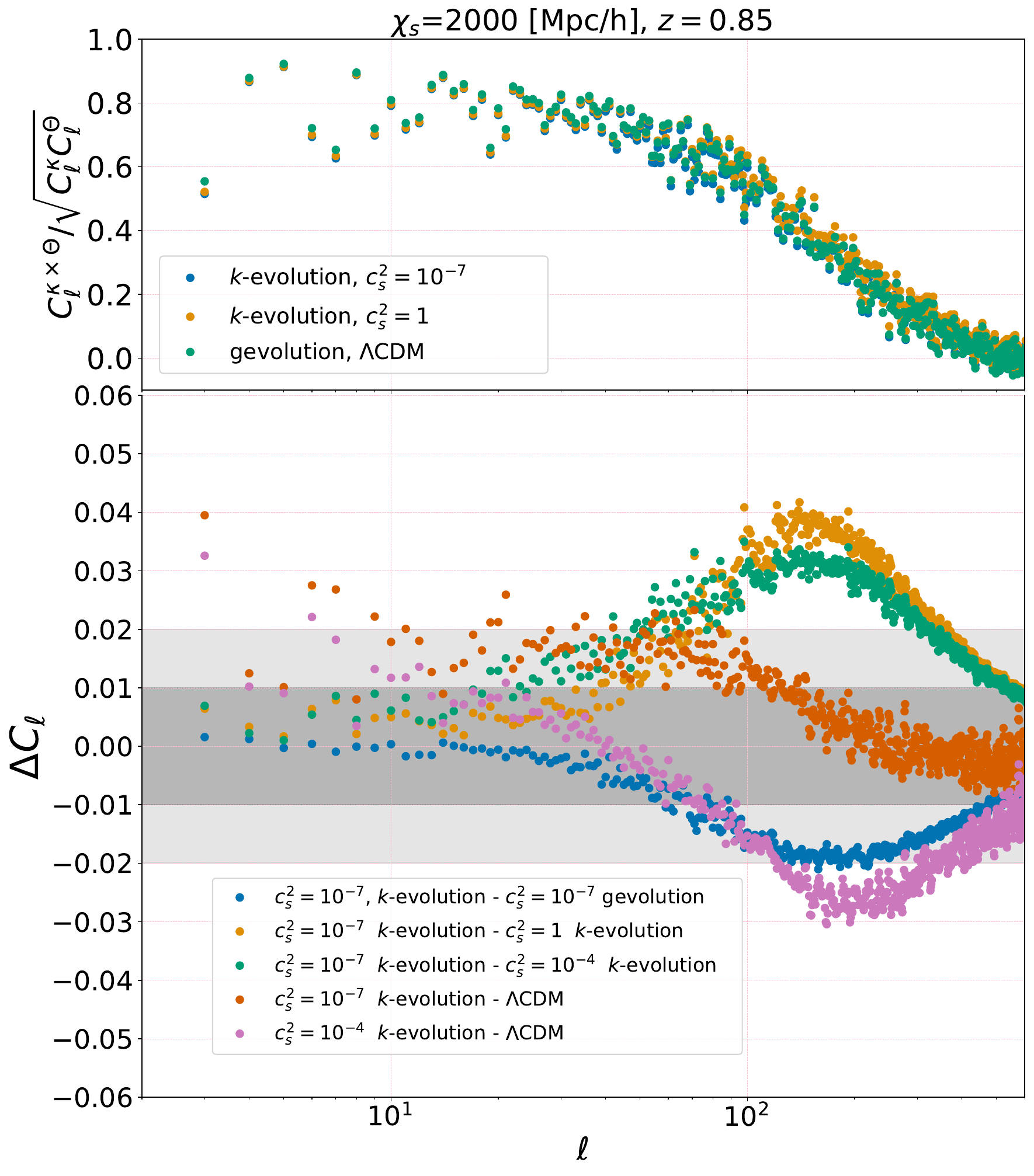} 
           \caption{\textbf{\emph{Top panel:}} The 
           {cross-correlation coefficient}
           of lensing convergence $\kappa$ and ISW-RS integrated to comoving distance ${\chi_s}=2000$ Mpc$/h$ {for the results shown in Fig.~\ref{fig:ISW-kappa_2000}. \textbf{\emph{Bottom panel:}} Absolute difference in the cross-correlation coefficient for different models.}} 
\label{fig:ISW-kappa_alm_2000}
\end{figure}

\subsection{Gravitational redshift}
In Fig.~\ref{fig:GR_redshift} the gravitational redshift {angular} power spectra for different scenarios are compared. {The top panel shows} $\ell (\ell + 1) C_{\ell}^{\delta z_\text{grav}}$ for three cases, and the bottom {panel} the relative difference between different angular power spectra. Like {for the} lensing signal, the effect of clustering and background almost cancel at large scales by coincidence: Comparing power spectra from dark energy with two different speeds of sound ($c_s^2=1$ and $c_s^2=10^{-7}$) {we find a difference} of $\sim 2\%$ due to the clustering of dark energy. The relative difference between $\Lambda$CDM with $w$CDM with $c_s^2=1$ reaches $\sim -3\%$ due to different background evolution. {But the difference between $\Lambda$CDM and $k$-essence with a low sound speed is only about 1\%.}
{Fig.~\ref{fig:GR_redshift} also shows that the effect of non-linear dark energy clustering becomes visible once again for $\ell\gtrsim100$, and is generally rather small, of the order of 1\%.}

{It should be noted that the angular correlation of the gravitational redshift will be very difficult to measure from large-scale structure observations. However, the line-of-sight correlations produce interesting signatures in redshift space, in particular a dipole in the correlation function of different matter tracers \citep{Wojtak:2011ia,Breton:2018wzk}.}

\begin{figure}
    \includegraphics[width=\columnwidth]{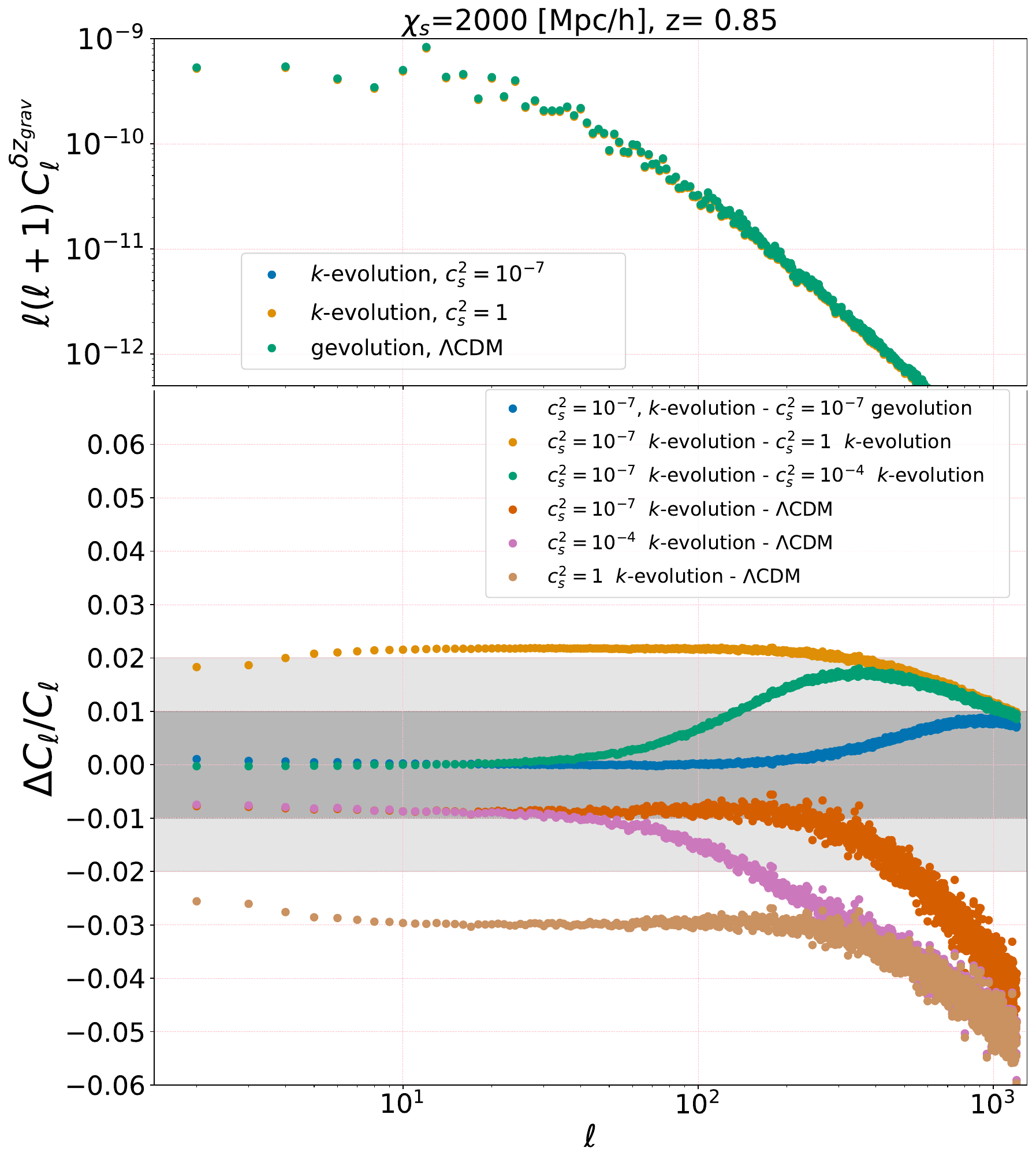}
           \caption{\textbf{\emph{Top panel:}} The gravitational {redshift} angular power spectra {of} different cosmologies for a source {plane} at comoving distance ${\chi_s}=2000$ Mpc$/h$. 
           \textbf{\emph{Bottom panel:}} The relative difference between the {angular} power spectra 
           {of the gravitational redshift in different cosmologies.}}
\label{fig:GR_redshift}
\end{figure}

\subsection{Shapiro time delay}
In the Fig.~\ref{fig:shapiro_2000} the angular power spectra for the Shapiro delay signal is plotted. 
{Our result for the relative spectra show a similar pattern as for the gravitational redshift. 
The effect of dark energy clustering and dark energy background evolution have a $\sim 3\%$ effect on the Shapiro delay power spectrum that again cancel to a significant degree when combined. The non-linear dark energy clustering contributes again about a 1\% effect for $\ell\gtrsim100$.} We only report {the} Shapiro delay integrating to {$\chi_s = 2000$ Mpc$/h$. The cosmological Shapiro time delay will be extremely challenging to measure. It contributes only to subdominant relativistic corrections in the redshift-space clustering, and we are not aware of a probe that easily isolates this effect.}

\begin{figure}
    \includegraphics[width=\columnwidth]{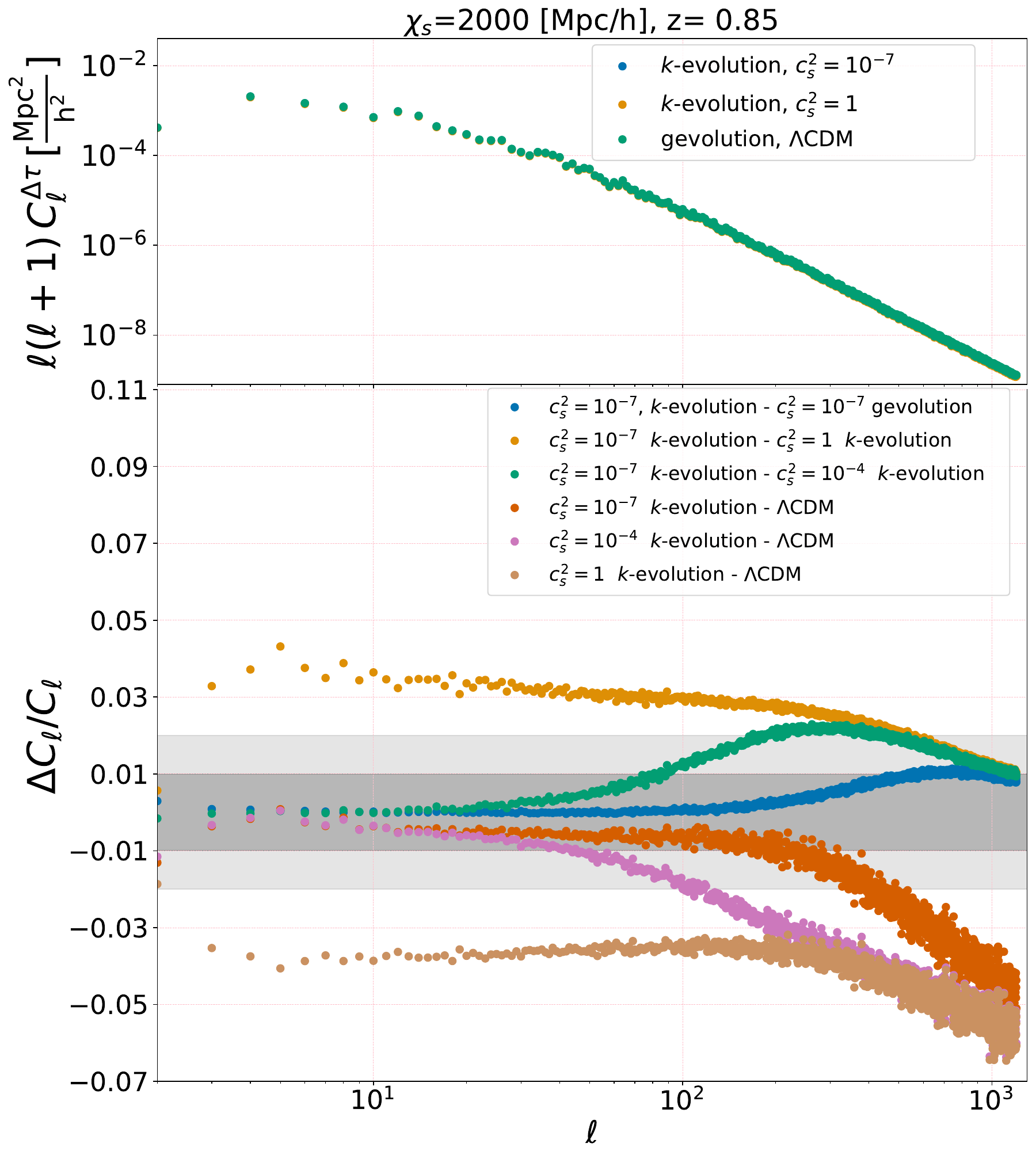}
           \caption{\textbf{\emph{Top panel:}} The angular power spectra of Shapiro time delay integrated to comoving distance ${\chi_s}=2000$ Mpc$/h$ {for different cosmologies.} 
           \textbf{\emph{Bottom panel:}} The relative difference between Shapiro delay {angular} power spectra for different models.} 
    \label{fig:shapiro_2000}
\end{figure}

\section{Conclusion} \label{sec:conclusion}
We are, observationally speaking, in the golden era of cosmology and in the near future we will be able to put stringent constraints on dark energy and modified gravity models. However, to be able to {unlock} the full power of future observations we need to have a precise 
{understanding of} the non-standard scenarios {well in}to the nonlinear regime. The precise modeling of structure formation and cosmological observables covering linear to nonlinear scales {can be directly achieved} using {full} $N$-body simulations.

In this work we have presented the numerical results for the effect of clustering dark energy ({specifically the} \kess\ model) on the observables extracted from the gravitational potential, namely the weak gravitational lensing, ISW-RS, Shapiro time delay and gravitational redshift. The observables discussed in this paper are {calculated} via {a} ray-tracing method  
integrating to the source redshifts $z\approx0.85$ and $z\approx 3.3$ {covering} respectively a full sky map and a pencil beam {in our simulations}. Comparing {results from} the two $N$-body codes \gev\ and \kev\ with the linear Boltzmann code \class\ 
we are able to {assess} the different effects coming from the dark energy, specifically the effect from a different background evolution, {from} linear dark energy perturbations and {from} the non-linear evolution of dark energy {itself}.

In sum{mary}, our numerical analysis of the {angular} power spectra of each observable shows that:
\begin{itemize}
  \item  The ISW-RS signal is the most powerful probe of different background evolution and the clustering of dark energy. Comparing dark energy with {various} speeds of sound ($c_s^2 = 10^{-7}$, $c_s^2 = 10^{-4}$ and $c_s^2 = 1$) with $\Lambda$CDM one finds a {significant} impact of clustering dark energy on the ISW-RS {angular} power spectra reaching $\sim 35\%$ at $z = 3.3$ and $\sim 30\%$ at $z = 0.85$. Moreover, by comparing the linear ISW signal from \class\ with {the} non-linear signal from our $N$-body codes we are able to {determine} the scale and amplitude of the nonlinear Rees-Sciama effect in the \kess\ scenario.
 \item  The effect of dark energy on the weak gravitational lensing signal could reach $\sim 5 \%$. {Interestingly,} our numerical study shows that the effect of clustering of dark energy and background evolution {can} partially cancel each other.
 \item The gravitational redshift and Shapiro time delay signals are less sensitive to the dark energy clustering and background evolution, as dark energy would change these signals at most {by} $\sim 3\%$ and $\sim 4\%$, respectively, due to the different background evolution. {An additional change by} $\sim 2\%$ and $\sim 3\%$ {can occur} due to the clustering of dark energy. 
\end{itemize}
{Our numerical study shows how direct probes of the gravitational field, in particular weak lensing and the ISW-RS effect, can be used to constrain the nature of dark energy. It also highlights the relevance of including non-linear effects and provides a framework to model these effects in full $N$-body simulations. With this we deliver some valuable guidance for the implementation of analysis pipelines that will be used to process the vast amount of upcoming observational data, with the aim to derive robust constraints on dark energy parameters.}

\section*{Acknowledgements}
FH would like to thank Francesca Lepori for helping in the maps analysis, Mona Jalilvand for helping in making linear angular power spectra and Shant Baghram for his comments about cross-correlating LSS data with ISW-RS signal. FH also would like to appreciate invaluable support from Jean-Pierre Eckmann during COVID-19 outbreak and his comments about our manuscript. {FH and MK acknowledge funding from the Swiss National Science Foundation, and} {JA acknowledges funding by STFC Consolidated Grant ST/P000592/1. This work was supported by a grant from the Swiss National Supercomputing Centre (CSCS) under project ID s969.} {Part of the computations were performed at University of Geneva on the Baobab cluster.}

\paragraph*{Carbon footprint:} {Our simulations}
 consumed about 
$9800 \, \mathrm{kWh}$ of electrical energy, which {is equivalent to $1960\, \mathrm{kg}\, \mathrm{CO}_2$ with a} conversion factor of $0.2\, \mathrm{kg}\,\mathrm{CO}_2\,\mathrm{kWh}^{-1}$ from \cite{VUARNOZ2018573}, table 2, assuming Swiss mix.

\section*{Data availability}
{
The
$N$-body codes used for the simulations and data of
this paper will be provided upon request to the corresponding author.} {The version of \gev\ used in this work, together with the map-making tools, is available on this public repository: \url{https://github.com/gevolution-code/gevolution-1.2}}

\appendix

\section{Convergence test}
\label{app:convergence}
In this appendix we compare the results from {our} high-resolution simulations ($N_\text{grid} = N_\text{pcl} = 4608^3$) with {some} lower{-resolution} ones ($N_\text{grid} = N_\text{pcl} = 2304^3$) for ISW-RS and convergence maps at both redshifts $z=0.85$ and $z=3.3$. 
{This shows us at which scales the low-resolution simulations have converged to a certain numerical precision, and we can make a reasonable guess on how well we should be able to trust our final results from the high-resolution simulations.}

According to Fig.~\ref{fig:convergence_kappa} the convergence angular power spectrum is converged to within {$5\%$} up to $\ell \sim 100$ and $\ell \sim 250$, respectively for a source redshift of $z = 0.85$ and $z = 3.3$. Assuming that a factor of $2$ in spatial resolution translates to a similar improvement in the angular resolution, we may want to trust our final results up to $\ell \sim 200$ and $\ell \sim 500$, respectively, maintaining a $\sim {5}\%$ absolute convergence threshold. {However, t}he leading-order error due to finite resolution is usually almost independent of cosmology, which means that it cancels out to high accuracy when taking ratios. Whenever we show relative changes, we therefore expect much better numerical convergence. {In the bottom panel of the Fig.~\ref{fig:convergence_kappa} we show the relative change between $\Lambda$CDM and $w$CDM convergence power spectra for two different spatial resolutions. {In these relative spectra the curves with different resolutions agree at all $\ell$ shown, explicitly demonstrating} the cancellation of the finite resolution error in the relative power spectra.} 

Considering Fig.~\ref{fig:convergence_ISW} we can draw similar conclusions for the numerical convergence of the ISW-RS signal. The ${5}\%$ error is reached at somewhat higher $\ell$, which means we may be able to trust our final results up to $\ell \sim {400}$ and $\ell \sim {800}$ for the two redshift values if we apply the same requirement on absolute numerical convergence. {Also according to the bottom panel of the figure the relative spectra from the two different resolutions are {consistent} so that we can trust the ISW-RS relative difference power spectra at all scales of interest.}
{We emphasise again} that the results shown in this paper were obtained using the higher-resolution simulations.

\begin{figure}
    \includegraphics[width=\columnwidth]{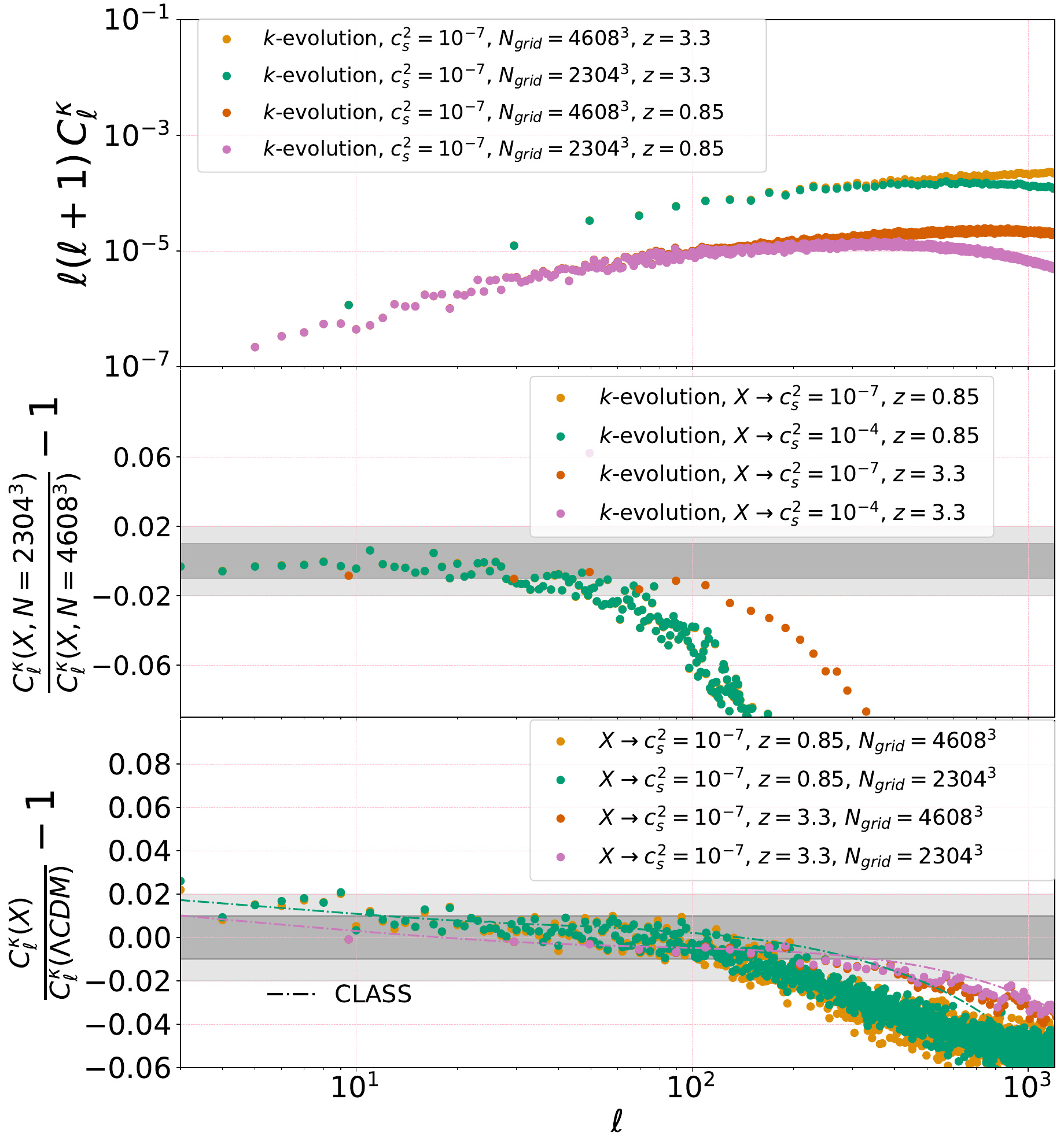}
    \caption{\textbf{\emph{Top panel:}} 
    {We compare} {the convergence power spectra from two simulations with different resolutions, $N_\text{grid} = N_\text{pcl} = 4608^3$ and $N_\text{grid} = N_\text{pcl} = 2304^3$, at two source redshifts $z = 0.85$ and $z = 3.3$. \textbf{\emph{Middle panel:}} The relative difference between the convergence power spectra of the same cosmology but different resolutions {of the simulations} are shown. This gives us {an} estimation of {the} finite resolution error on {the} convergence {angular} power spectra. {The grey areas show {$1\%$ and $2\%$} numerical agreement.} 
     \textbf{\emph{Bottom panel:}} Comparing the relative change in the convergence power spectra between $\Lambda$CDM and $k$-essence cosmology with the speed of sound $c_s^2 =10^{-7}$  from two simulations with different resolutions, $N_\text{grid} = N_\text{pcl} = 4608^3$ and $N_\text{grid} = N_\text{pcl} = 2304^3$, at source redshifts $z = 0.85$ and $z = 3.3$. {The dashed lines show the linear theory prediction obtained from \class.} {The agreement across different resolutions shows that} one can {trust} the relative change of power spectra at much higher {multipoles than is the case for} the {individual} power spectra themselves.} }
       \label{fig:convergence_kappa}
\end{figure}
\begin{figure}
    \includegraphics[width=\columnwidth]{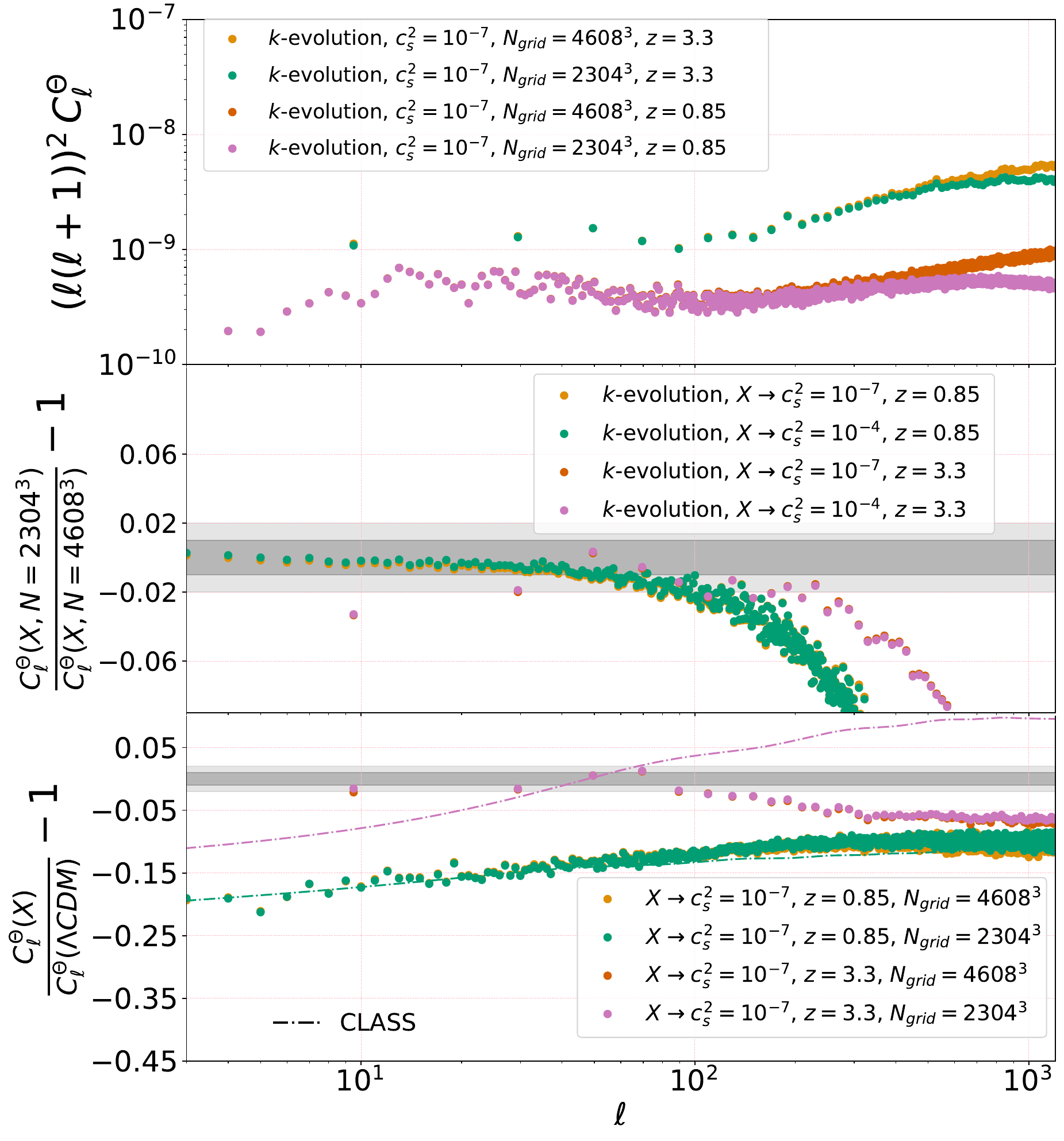} 
    \caption{\textbf{\emph{Top panel:}} {We compare} {the ISW-RS {angular} power spectra from two simulations with different resolution, $N_\text{grid} = N_\text{pcl} = 4608^3$ and $N_\text{grid} = N_\text{pcl} = 2304^3$, at two source redshifts $z=0.85$ and $z=3.3$. \textbf{\emph{Middle panel:}} The relative difference between the ISW-RS {angular} power spectra of the same cosmology but different resolutions {of the simulations} are shown.  \textbf{\emph{Bottom panel:}} The relative difference of ISW-RS {angular} power spectra between the $k$-essence cosmology with $c_s^2 =10^{-7}$ and $\Lambda$CDM for two different spatial resolutions  $N_\text{grid} = N_\text{pcl} = 4608^3$ and $N_\text{grid} = N_\text{pcl} = 2304^3$ at two source redshifts $z = 0.85$ and $z = 3.3$. {The dashed lines show the linear theory prediction obtained from \class.} {Like Fig.~\ref{fig:convergence_kappa}, t}his figure shows that the finite resolution effect is cancelled significantly 
    in the relative changes of the angular power spectra.
    }
    } 
    \label{fig:convergence_ISW}
\end{figure}

\section{Non-linear Rees-Sciama effect}
\label{app:NLRS}

In this appendix we {quantify in more detail the contribution of the non-linear Rees-Sciama effect to the ISW-RS signal at low multipoles $\ell \lesssim 100$. To this end, we run additional simulations for the $w$CDM cosmology} 
with $c_s^2=1$ and {for} $\Lambda$CDM{, but with the power of primordial perturbations, $A_s$, reduced by two orders of magnitude. As a result, the evolution is almost linear up to much smaller scales, bringing the numerical result for the ISW-RS signal into very good agreement with the calculation of the linear ISW alone as performed by \class. This can be clearly seen in the left panel of Fig.~\ref{fig:RS-test} where the simulations with low $A_s$ remain consistent with the prediction from \class\ well beyond $\ell \sim 100$. Taking the ratio of spectra from the simulations with the standard value of $A_s$ to the ones with low $A_s$, we can therefore get a good estimate of the fractional contribution of the Rees-Sciama effect.}
{The result of this analysis is shown in the right panel of Fig.~\ref{fig:RS-test}, which suggests that the Rees-Sciama effect changes the signal by a few per cent at very low $\ell \lesssim 20$, and then gradually ramps up to reach a $\sim 100 \%$ correction at around $\ell \sim 100$. Interestingly, the non-linear corrections are systematically larger in $\Lambda$CDM, which can be explained by the fact that the linear growth rate is slightly suppressed in our $w$CDM cosmology, see Fig.~\ref{fig:BG}.} {This, in turn, is expected to lead to a corresponding relative suppression of second-order corrections like the Rees-Sciama effect.}

\begin{figure*}
\begin{minipage}{\textwidth}
\centering
\includegraphics[width=.49\textwidth]{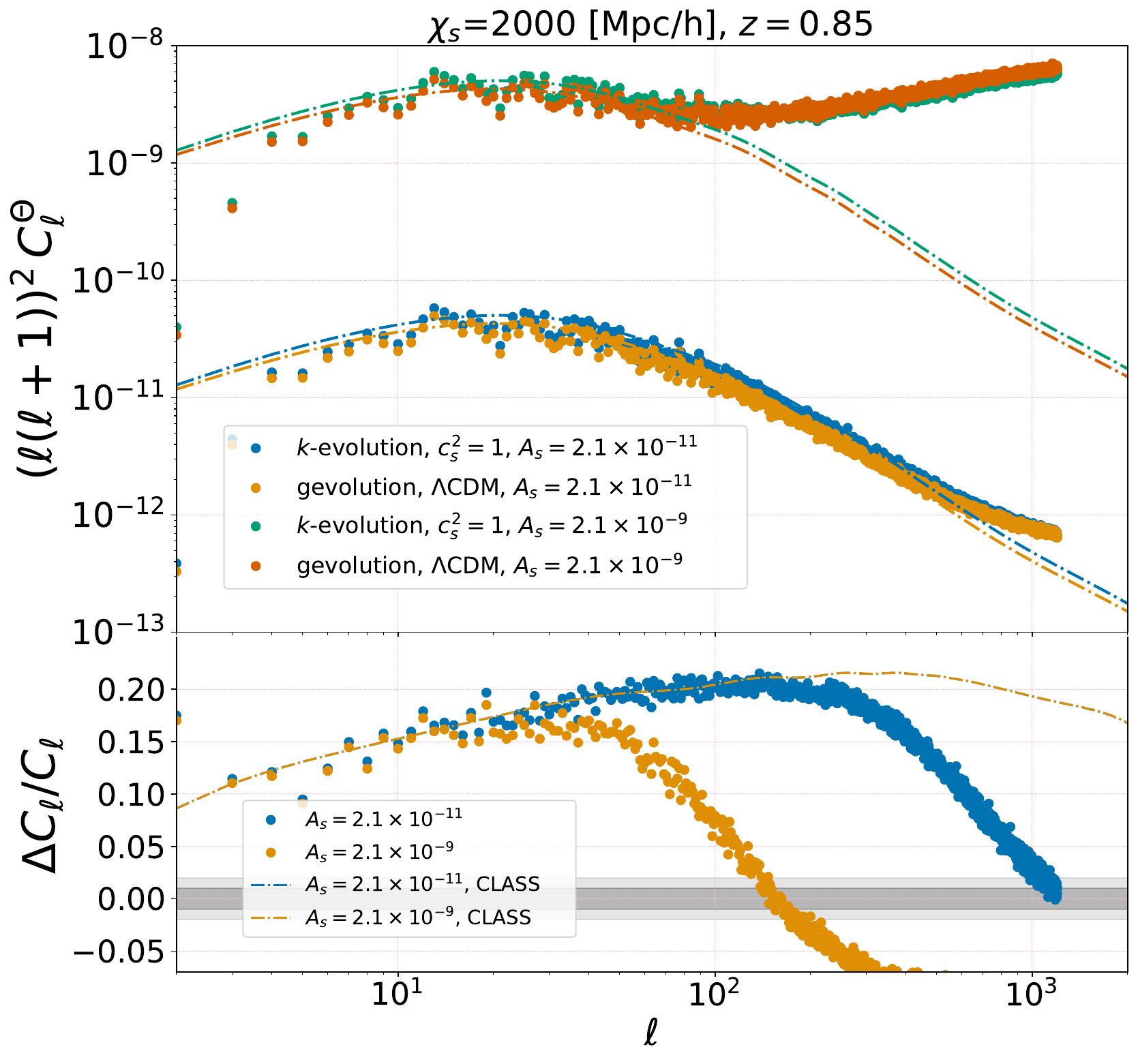}\hfill
\includegraphics[width=.49\textwidth]{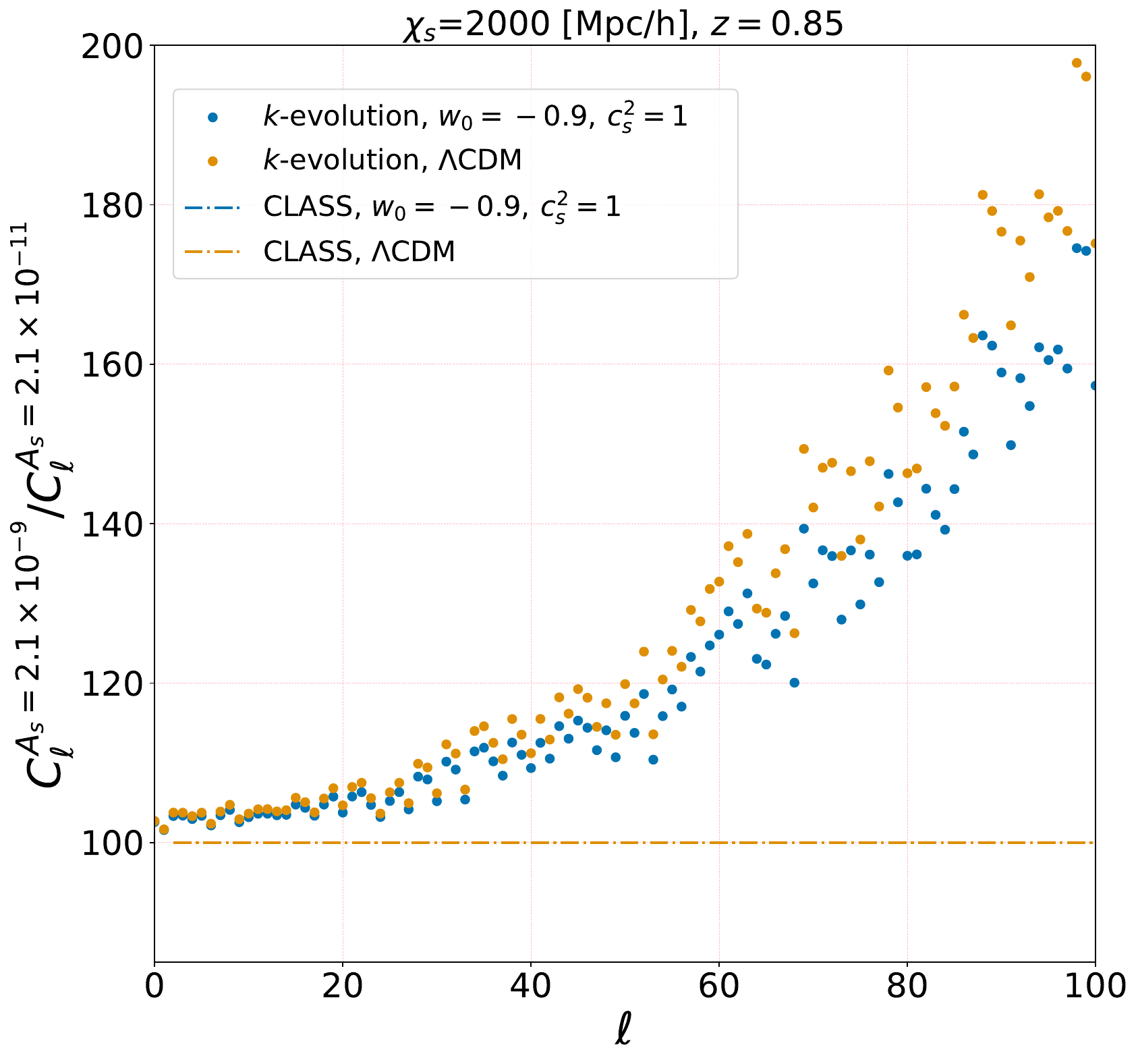}\hfill
\caption{\textbf{\emph{Left:}} The ISW-RS angular power spectra for two different cosmologies, ($w_0=-0.9$, $c_s^2=1$) and $\Lambda$CDM with {standard} and {low} value of $A_s$, the amplitude of scalar perturbations.
\textbf{\emph{Right:}} The ratio of the ISW-RS {signal} from high value {of} $A_s$ to low value of $A_s$ are compared. The linear result is constant at all scales{, and is simply the ratio of the two values of $A_s$.} 
{T}he fractional contribution of the Rees-Sciama effect depends on $w_0$ mainly because {of suppressed growth and the corresponding shift of} the non-linear scale 
with respect to $\Lambda$CDM.}
\label{fig:RS-test}
 \end{minipage}
 \end{figure*}
 
 \bibliographystyle{mnras}
\bibliography{inspire}

\end{document}